\documentclass[fleqn,usenatbib]{mnras}

\usepackage{newtxtext,newtxmath}
\usepackage{xcolor}
\usepackage[T1]{fontenc}

\DeclareRobustCommand{\VAN}[3]{#2}
\let\VANthebibliography\thebibliography
\def\thebibliography{\DeclareRobustCommand{\VAN}[3]{##3}\VANthebibliography}

\usepackage{graphicx}	% Including figure files
\usepackage{amsmath}	% Advanced maths commands
\title[Prospects for Obs GWs from Fast-spinning WDs]{Prospects for the Observation of Continuous Gravitational Waves from Deformed Fast-spinning White Dwarfs}

\author[M. F. Sousa et al.]{
Manoel F. Sousa,$^{1,2}$\thanks{E-mail: manoelsousa@utfpr.edu.br}
Edson Otoniel,$^{3}$
Jaziel G. Coelho$^{2,4}$
and Jos\'e C. N. de Araujo$^{2}$
\\
$^{1}$Programa de Pós-Graduação em Física e Astronomia, Universidade Tecnológica Federal do Paraná, Avenida sete de setembro 3165, 80230-901, Curitiba, PR, Brazil\\
$^{2}$ Divis\~{a}o de Astrof\'{i}sica, Instituto Nacional de Pesquisas Espaciais, Avenida dos Astronautas 1758, S\~{a}o Jos\'{e} dos Campos, SP 12227-010, Brazil \\
$^{3}$Instituto de Formaç\~{a}o de Educadores, Universidade Federal do Cariri, R. Oleg\'{a}rio Emidio de Araujo, s/n – Aldeota, 63260-000 Brejo Santo, CE, Brazil\\
$^{4}$N\'ucleo de Astrof\'{\i}sica e Cosmologia (Cosmo-Ufes) \& Departamento de F\'isica, Universidade Federal do Esp\'irito Santo, 29075--910, Vit\'oria, ES, Brazil
}

\date{Accepted XXX. Received YYY; in original form ZZZ}

\pubyear{2023}

\begin{document}
\label{firstpage}
\pagerange{\pageref{firstpage}--\pageref{lastpage}}
\maketitle

\begin{abstract}
There has been a growing interest within the astrophysics community in highly magnetized and fast-spinning white dwarfs (WDs), commonly referred to as HMWDs. WDs with these characteristics are quite uncommon and possess magnetic fields $\geqslant 10^6$~G, along with short rotation periods ranging from seconds to just a few minutes. Based on our previous work, we analyze the emission of Gravitational Waves (GWs) in HMWDs through two mechanisms: matter accretion and magnetic deformation, which arise due to the asymmetry surrounding the star’s rotational axis. Here, we perform a thorough self-consistent analysis, accounting for rotation and employing a realistic equation of state to investigate the stability of stars. Our investigation focuses on the emission of gravitational radiation from six rapidly spinning WDs: five of them are situated within binary systems, while one is an AXP, proposed as a magnetic accreting WD. Furthermore, we apply the matter accretion mechanism alongside the magnetic deformation mechanism to assess the influence of one process on the other. Our discoveries indicate that these WDs could potentially act as GW sources for BBO and DECIGO, depending on specific parameters, such as their mass, the angle ($\alpha$) between the magnetic and rotational axes, and the accumulated mass ($\delta m$) at their magnetic poles, which is influenced by the effect of matter accretion. However, detecting this particular class of stars using the LISA and TianQin space detectors seems unlikely due to the challenging combination of parameters such as a large $\delta m$, a large $\alpha$ angle and a small WD mass value.

\end{abstract}


%%%%%%%%%%%%%%%%%%%%%%%%%%%%%%%%%%%%%%%%%%%%%%%%%%

%%%%%%%%%%%%%%%%% BODY OF PAPER %%%%%%%%%%%%%%%%%%

%%%%%%%%%%%%%%%%%%%%%%%%%%%%%%%%%%%%%%%%%%%%%%%%%%%%%%%%%%%%%%%
%%%%%%%%%%%%%%%%%%%%%%%%%%%%%%%%%%%%%%%%%%%%%%%%%%%%%%%%%%%%%%%
\section{Introduction}
\label{sec:intro}
%%%%%%%%%%%%%%%%%%%%%%%%%%%%%%%%%%%%%%%%%%%%%%%%%%%%%%%%%%%%%%%
%%%%%%%%%%%%%%%%%%%%%%%%%%%%%%%%%%%%%%%%%%%%%%%%%%%%%%%%%%%%%%%
White dwarfs (WDs) represent the ultimate stage of stellar evolution for stars with masses ranging from $\sim0.4~\textrm{M}{\odot}$ to $8.0~\textrm{M}{\odot}$. Some of these remnants rotate with periods ranging from seconds to years. In general, isolated WDs have rotation periods of days or even years. However, isolated WDs rotating with periods of a few seconds to minutes have been observed recently; as examples, we can mention SDSS~J221141.80+113604.4 (SDSS~J2211+1136) with a rotation period of $P = 70.32$~s \citep{2021ApJ...923L...6K} and ZTF~J190132.9+145808.7 (ZTF~J1901+1458) with $P = 416.20$~s \citep{2021Natur.595...39C}. WDs in binary systems, on the other hand, spin with periods ranging from seconds to hours.
Recently, \cite{Pelisoli2022} reported on a WD exhibiting the most fast rotation ever observed. In their study, the authors present observations conducted with the  \textit{Gran Telescopio Canarias} (GTC), unveiling that LAMOST J0240.51+1952 is a cataclysmic variable hosting a WD with an extremely fast rotation of $24.93$~s. 
Another WD with a very similar rotation period was detected in the binary system CTCV~J2056-3014 from the XMM-Newton observations that point out that the WD has a rotation period of $29.6$~s \citep{LopesDeOliveira2020}.

From a theoretical perspective, WDs can have an extremely short rotation period and still maintain rotational stability. For instance, \cite{2013ApJ...762..117B} investigated the stability of WDs for different nuclear compositions using General Relativity to describe uniformly rotating WDs within Hartle's formalism \citep{1967ApJ...150.1005H,1968ApJ...153..807H}. In this scenario, they found that the minimum rotation periods for a WD composed of $^{4}$He, $^{12}$C, $^{16}$O and $^{56}$Fe be stable are $P_{min} \sim$ ($0.3$; $0.5$; $0.7$; $2.2$)~s with a maximum mass of $M \sim$ ($1.500$; $1.474$; $1.467$; $1.202$)~$\textrm{M}_\odot$, respectively.

WDs can also have magnetic fields with intensities ranging from $10^3$~G to $10^9$~G \citep{2015SSRv..191..111F}. In recent years, there has been an increasing interest within the astrophysics community in highly magnetized white dwarfs (HMWDs). These HMWDs, with surface magnetic fields ranging from $10^6$~G to $10^9$~G, have already been confirmed in several works \cite[see e.g.,][]{2009A&A...506.1341K, 2010yCat..35061341K, kepler/2013, 2015MNRAS.446.4078K, 2023ApJ...944...56A}. In addition to their high magnetic fields, most of these WDs are massive; for instance: REJ 0317--853 has $M \approx 1.35~\textrm{M}_\odot$ and $B\approx (1.7$--$6.6)\times 10^8$~G \citep{1995MNRAS.277..971B,2010A&A...524A..36K}; ZTF~J1901+1458 has $M \approx 1.31~\textrm{M}_\odot$ and $B\approx (6.0$--$9.0)\times 10^8$~G \citep{2021Natur.595...39C}; and PG 1031+234 has the highest magnetic field $B\approx 10^9$~G \citep{1986ApJ...309..218S,2009A&A...506.1341K}. The existence of ultra-massive WDs has been revealed in several studies~\cite[see e.g.,][]{2005A&A...441..689A,2007A&A...465..249A,2013ApJ...771L...2H,2017MNRAS.468..239C,2019A&A...625A..87C,10.1093/mnras/sty3016,2018MNRAS.480.4505J}.

Some rapidly rotating magnetic WDs have been observed to emit pulsed electromagnetic radiation and have been proposed as WD pulsars. In particular, a WD pulsar was recently discovered in a binary system known as AR~Scorpii (AR~Sco). This WD has a rotation period of $1.97$~min and emits pulses over a wide range of frequencies \citep{2016Natur.537..374M}. The spindown luminosity of this WD is an order of magnitude greater than the observed X-ray luminosity, which, with the absence of accretion signals, suggests that AR~Sco is probably a rotation-powered pulsar \citep{2017NatAs...1E..29B}. More recently, another WD pulsar, considered to be a sibling of AR~Sco, has been discovered in the system J191213.72$-$441045.1 \citep{2023NatAs.tmp..120P}. This pulsar rotates with a period of $5.30$~min and exhibits pulsed emission ranging from radio to X-ray wavelengths. The spindown value for this source has not yet been measured. With that, constraints on the strength of the magnetic field and the mechanism that powers the pulsar have not been estimated \citep{2023NatAs.tmp..120P}.

Other sources have also been proposed as WD pulsar candidates. A specific example is AE~Aquarii (AE~Aqr), identified as an X-ray pulsar with a short rotation period of $P = 33.08$~s \citep{1979ApJ...234..978P, Terada2008}. Another example is the HD~49798/RX~J0648.0-4418 binary system. \cite{Mereghetti2009} showed that the pulsed X-ray emission from this system came from a massive WD with mass $M=1.28M_\odot$ and a very short rotation period of $P = 13.2$~s. However, the nature of this compact object is still unclear \cite[see][]{2016MNRAS.458.3523M, 2018MNRAS.474.2750P}.

Some isolated, very massive, magnetic, and fast WDs seem to have been originated by a merger of two WDs. Numerical simulations show that these WDs might indeed originate in double WD mergers \cite[see e.g.,][]{2009A&A...500.1193L,2014MNRAS.438...14D}. The general configuration after the merger is a central WD formed by the mass of the primary WD, which is rigidly rotating. This central object is surrounded by a hot corona with differential rotation and a Keplerian disc of rapidly rotating mass. The mass of the interrupted secondary star is distributed between the envelope and the disk. The hot, convective corona works as an efficient dynamo that can lead to magnetic fields of $\lesssim 10^{10}$~G \citep{2012ApJ...749...25G}. From this post-merger configuration, \cite{2022ApJ...941...28S} recently investigated the possibility that the isolated WDs, SDSS~J2211+1136 and ZTF~J1901+1458, are products of the coalescence of double WDs. For this, the authors calculated the post-merger rotational evolution of the central WD and inferred the system parameters for which the rotational age coincides with the cooling age of the WD. They concluded that the observed parameters of SDSS~J2211+1136 and ZTF~J1901+1458 are consistent with an origin in the double WD mergers. More recently, \cite{2023ApJ...958..134S} characterized the optical signatures of these mergers. They found that the merger ejecta's expansion and cooling produce a peak optical emission $1$--$10$~days post-merger, with luminosities of $10^{40}$--$10^{41}$~erg~s$^{-1}$ and that LSST at the Vera C. Rubin Observatory will likely detect a few hundred of these events annually.

Several theoretical works regarding very magnetic, massive, and fast WDs have been published in the last few years. One of these studies, in particular, involves WD pulsars in an alternative description for the Soft Gamma Repeater (SGRs) and Anomalous X-Ray Pulsars (AXPs) \citep{2012PASJ...64...56M, 2014PASJ...66...14C, 2016JCAP...05..007M}. From this perspective, a canonical model of a pulsar powered by rotation can explain the process of energy emission in a WD \citep{1988PAZh...14..606U, 2017A&A...599A..87C}. In another proposed scenario, \cite{2013ApJ...772L..24R} suggest that AXP~4U~0142+61 is a WD pulsar resulting from the merger of two common WDs, which is surrounded by an accretion disk produced during the merger. The optical and infrared observational data from this source would thus be explained by the WD photosphere and the disk, while the X-ray emission would be caused by a pulsar-like emission. However, \cite{Borges2020} argue that, within this perspective, quiescent spectral energy distribution from mid-infrared down to hard X-rays can be described by a magnetic and isolated WD that accretes matter from a debris disk that surrounds it.

From a gravitational perspective, direct observations of gravitational waves (GWs) have recently been made by the LIGO-Virgo-KAGRA collaboration. The first event was detected in 2015 by the LIGO detector \citep{2016PhRvL.116f1102A}. Since then, a few dozen detections of GWs have been made that come from the coalescence of compact binaries, namely black hole binaries, neutron star binaries, and binaries formed by a black hole and a neutron star \citep{2019PhRvX...9c1040A,2021PhRvX..11b1053A,2021arXiv211103606T}. All these GWs detections are in a frequency band ranging from $10$~Hz to $1000$~Hz, which is the operational band of LIGO, Virgo, and KAGRA. However, there are several proposals for space missions for detection at lower frequencies, such as LISA and TianQin, whose frequency band is of $(10^{-4}-10^{-1})$~Hz, and BBO and DECIGO over a frequency band ranging from $0.01$~Hz to $10$~Hz \cite[see e.g.,][]{AMARO/2017, 2016CQGra..33c5010L, 2006CQGra..23.4887H, 2017PhRvD..95j9901Y, 2006CQGra..23S.125K}.

Different  possibilities  of  generation  of continuous GWs have already been proposed~\citep[see e.g.,][and references therein]{1996A&A...312..675B,2016ApJ...831...35D,2017MNRAS.472.3564M,2017EPJC...77..350D,2017ApJ...844..112G,2018EPJC...78..361P,2020ApJ...896...69K}. \cite{2017MNRAS.467.4484F} investigated the possibility that fast-spinning and magnetic WDs are sources of detectable GW emission. They computed the quadrupole moment of the mass distribution and performed an estimate of the GW of such WDs from the construction of numerical stellar models at different baryon masses. They found that magnetic and fast WDs might generate GW radiation that lies in the bandwidth of BBO and DECIGO. \cite{2019MNRAS.490.2692K} also showed that fast and magnetic WDs can be prominent sources for space detectors such as DECIGO, BBO, and LISA. The authors explored very massive WDs (close to the Chandrasekhar limit) and super-Chandrasekhar WDs and analyzed the GW amplitude for different WD parameters. They found that several super-Chandrasekhars, rotating in a frequency range of $0.01 - 1$~Hz and with a distance of $100$~pc, emit gravitational radiation detectable by BBO, DECIGO and LISA, considering a 1-year observation \citep[see also][]{2020ApJ...896...69K, 2022Parti...5..493M}.

The main goal of this paper is to extend our previous study \citep{2020MNRAS.492.5949S} in which we investigated the continuous gravitational radiation from three fast-spinning magnetized WDs, which rotate with periods from seconds to minutes and have magnetic fields around $10^{6}$~G to $\sim 10^{9}$~G. We considered two mechanisms of emission of GWs: matter accretion and magnetic deformation. In both cases, WDs generate continuous GWs due to the asymmetry around their axis of rotation formed as a result of the accumulated mass at the magnetic poles of a WD accreting mass, so that the accretion flux follows the field lines, or due to the intense magnetic field, which can break the star's spherical symmetry, making it oblate or prolate depending on the field configuration. Here, we increase the number of fast-spinning and magnetic WDs and investigate the combined impact of accretion and magnetic deformations on the emission of GWs. Specifically, we examine how the asymmetry induced in the WD by accretion and the magnetic field synergistically influence the emission of gravitational radiation. Additionally, we calculate the GW amplitude for a mass range derived from theoretical limits (minimum and maximum) obtained through numerical and self-consistent determination of stability regions for WDs. This determination incorporates constraints imposed by the Kepler frequency (mass-shedding) and pycnonuclear fusion reactions.

This paper is organized as follows. In Sec. \ref{sec:model}, we present the basic model adopted, brief descriptions of the Hartle's stellar rotation formalism and the pycnuclear fusion reactions. The gravitational emission mechanism is considered in Sec. \ref{sec:GW_emission}. In Sec. \ref{sec:res_dis} the results and discussions are considered. Finally, in Sec. \ref{sec:concl} the main conclusions are presented.

%%%%%%%%%%%%%%%%%%%%%%%%%%%%%%%%%%%%%%%%%%%%%%%%%%%%%%%%%%%%%%%
%%%%%%%%%%%%%%%%%%%%%%%%%%%%%%%%%%%%%%%%%%%%%%%%%%%%%%%%%%%%%%%
\section{The model} \label{sec1}
\label{sec:model}
%%%%%%%%%%%%%%%%%%%%%%%%%%%%%%%%%%%%%%%%%%%%%%%%%%%%%%%%%%%%%%%
%%%%%%%%%%%%%%%%%%%%%%%%%%%%%%%%%%%%%%%%%%%%%%%%%%%%%%%%%%%%%%%

Recent advancements in observation and detection techniques for WDs have sparked significant interest in theoretical studies of their structure and evolution. This is accompanied by sophisticated calculations of fermionic matter properties under extreme conditions \citep{eisenstein_catalog_2006,kepler/2013,2015MNRAS.446.4078K,2019A&ARv..27....7C}. In this context, the equation of state (EoS) for relativistic and degenerate WD matter plays a crucial role. The EoS accounts for the excess free energy of an ion-based one-component plasma (OCP) and is computed using the Helmholtz free energy \citep{2013ApJ...762..117B,2014ApJ...794...86C,chamel_maximum_2014,2018ApJ...857..134B,otoniel_strongly_2019, 2022atcc.book..121O}.

The Helmholtz free energy, denoted by $F$, is given by:
\begin{equation}
  F=F_{\mathrm{ id}}^{\mathrm{ ion}}+F_{\mathrm{ id}}^{(\rm e)}+F_{\rm e e}+F_{\rm i i}+F_{\rm i e} \, .
\label{eq:F}
\end{equation}

The free energy of a nonrelativistic classical gas, $F_{\mathrm{id}}^{\mathrm{ion}}$, is described by:
\begin{equation}
F_{\mathrm{id}}^{\mathrm{ion}}=N_{\rm j} k_{\rm B} T\left[\ln \left(n_{\rm j} \lambda_{\rm j}^{3} / g_{\rm j}\right)-1\right] .
\end{equation}

Herein, $N_{\rm j}=n_{\rm j} V$, $k_{\rm B}$ symbolizes the fundamental Boltzmann constant, $T$ corresponds to the temperature of the gas, $n_{\rm j}$ encapsulates the comprehensive number density of ions of an ensemble of particles in volume V and $\lambda_{\rm j}^{3}$ epitomizes the thermal de Broglie wavelength, whereas $\lambda_{\rm j}=\left(2 \pi \hbar^{2} / m_{\rm j} k_{\rm B} T\right)^{1 / 2}$ encapsulates the profound essence of the ion's mass $m_{\rm j}$ and the multiplicity of its spin, ingeniously denoted by $g_{\rm j}$. A meticulous examination of the seminal work by~\cite{PhysRevE.58.4941} allows us to glean a profound understanding of its intricate fabric.

Similarly, the free energy of electrons, $F_{\mathrm{id}}^{(\rm e)}$, can be expressed as:
\begin{equation}
F_{\mathrm{id}}^{(\rm e)}=\mu_{e} N_{e}-P_{\mathrm{id}}^{(\rm e)} V.
\end{equation}

The pressures and electron number density are given by:
$P_{\text{id}}^{(\rm e)}$ and $n_{\rm e}=N_{\rm e} / V$,  
where $N_e$ represents the number of electrons and $\mu_e$ denotes the electron chemical potential excluding the rest energy $m_e c^2$. This expression can be further elaborated in terms of Fermi-Dirac integrals, $I_{\rm \nu}\left(\chi_{\rm e}, \tau\right)$, where $\chi_{\rm e}=\mu_{\rm e}/k_{\rm B} T$ and $\nu=1 / 2,3 / 2,$ and $5 / 2 .$ The chemical potential can be obtained  by inverting the function $n_{\rm e}\left(\chi_{\rm e}, T\right)$ numerically. The remaining terms in equation (\ref{eq:F}) represent free energies arising from electron--electron, ion--ion, and ion--electron interactions \citep{PhysRevE.58.4941,potekhin_equation_2000}. The plasma's pressure ($P$) and entropy ($S$) with a fixed number of particles in a given volume ($V$) are obtained from the thermodynamic relations $P=-(\partial F / \partial V)_{T,N}$ and $S = -(\partial F  / \partial T)_{V,N}$, where the internal energy is defined as $U=F+T S$ \citep{PhysRevE.58.4941,potekhin_equation_2000}.

%%%%%%%%%%%%%%%%%%%%%%%%%%%%%%%%%%%%%%%%%%%%%%%%%%%%%%%%%%%%%%%%%%%%%%%
\subsection{Hartle's Stellar Rotation Formalism}\label{ssec:hartle}
%%%%%%%%%%%%%%%%%%%%%%%%%%%%%%%%%%%%%%%%%%%%%%%%%%%%%%%%%%%%%%%%%%%%%%%

The structural characteristics of WDs are governed by a state of hydrostatic equilibrium, wherein the gravitational forces are precisely counterbalanced by the outward pressure exerted by a relativistic electron gas. Recent studies have underscored the importance of accounting for the contributions of general relativity in accurately modeling the structural properties of WDs~\citep[see][]{boshkayev_maximum_2013}. The Tolman-Oppenheimer-Volkoff (TOV) equation serves as the cornerstone for describing the configuration of non-rotating stellar objects, providing crucial insights into their radii and masses. However, to fully incorporate the effects of rotation, it is imperative to solve Einstein's field equations using a metric that accommodates rotational deformations and the accompanying dragging of local inertial frames~\citep[see][]{1967ApJ...150.1005H,1968ApJ...153..807H,Friedman1986}.

In this study, we adopt Hartle's formalism to investigate the equilibrium configurations of rotating stellar bodies within the framework of general relativity. Hartle's formalism is founded upon two critical assumptions: uniform rotation and slow rotation, with a quadratic dependence on the angular velocity. These assumptions provide the necessary mathematical framework to model the behavior of rotating stars, enabling a systematic analysis of their structure and dynamics. By incorporating these assumptions, a deeper understanding of the physical processes that govern the behavior of rotating stellar objects is achieved. In this context, the metric formulation employed can be expressed as follows:

\begin{equation}
ds^2 = - e^{2 \nu} dt^2 + e^{2 \psi} (d\phi-\omega dt)^2 + e^{2\mu}
d\theta^2 + e^{2\lambda} dr^2 , 
\end{equation}
where the metric functions $\nu$, $\psi$, $\mu$, and $\lambda$, along with the frame dragging frequency $\omega$, are reliant upon the radial coordinate $r$, the polar angle $\theta$, and implicitly on the rotational frequency $\Omega$ of the star~\citep[see][for further details]{Friedman1986,weber_pulsars_1999}. Accordingly, we conducted a comprehensive two-dimensional analysis that incorporates the rotational deformations described by the polar angle $\theta$, while assuming axial symmetry in compact object configurations. The range of $\Omega$ considered was confined within $0 \leq \Omega \leq \Omega_K$, where $\Omega_K$ $(=2\pi/ P_{K})$ denotes the Kepler frequency, signifying the threshold for stable rotation termination due to mass-shedding. Remarkably, $\Omega_K$ serves as an absolute upper limit for rapid rotation. 

In our study of relativistic stars, we applied Hartle's perturbative formalism, which approximates the behavior of rotating stars up to the Keplerian frequency. While Hartle's work defines the slow rotation limit using $\sqrt{M/R^3}$ as a reference, it does not specify the upper limit of the rotation rate for perturbative calculations \citep{1967ApJ...150.1005H,1968ApJ...153..807H}. Nevertheless, \cite{1992ApJ...390..541W} have demonstrated through numerical analysis that Hartle's perturbative approach remains valid within a certain margin of deviation from exact results at the Keplerian frequency. Therefore, despite the Keplerian frequency being a characteristic frequency of the system, our justification for extending Hartle's formalism to encompass this frequency range lies in Weber $\&$ Glendenning's validation of its applicability beyond the traditional slow rotation limit.

The self-consistent computation of the Kepler frequency was performed concurrently with Einstein's field equations pertaining to the metric functions. This procedure entailed utilizing the following equation:

\begin{equation}
\Omega_{\rm K} = \omega + \frac{\omega'} {2\psi'} + e^{\nu-\psi}
\sqrt{\frac{\nu'}{\psi'} + \left(\frac{\omega'}{2\psi'} e^{\psi-\nu}
\right)^2} ,
\label{Kepler}
\end{equation}
where, the primes indicate partial derivatives of the metric functions with respect to the radial coordinate~\citep[][]{1992ApJ...390..541W,PhysRevD.50.3836}. Perturbative treatments, such as the one developed by Hartle, have been found to yield results that demonstrate remarkable agreement with those obtained through numerically exact approaches to Einstein's field equations. This notable congruity is particularly evident when examining the increase in mass attributable to rapid rotation at the Kepler frequency~\citep[][]{1992ApJ...390..541W}.

To summarize, a comprehensive understanding of WD structure necessitates the consideration of general relativity effects and the inclusion of rotational phenomena within appropriate metric formulations. These advancements provide invaluable insights into the interplay between gravitational forces, pressure, and rotation, thereby enriching our knowledge of these intriguing remnants of stellar evolution~\citep[][]{1968ApJ...153..807H}.

%%%%%%%%%%%%%%%%%%%%%%%%%%%%%%%%%%%%%%%%%%%%%%%%%%%%%%%%%%%%%%%%%%%%%%%
\subsection{Pycnonuclear fusion reactions}\label{ssec:pycno}
%%%%%%%%%%%%%%%%%%%%%%%%%%%%%%%%%%%%%%%%%%%%%%%%%%%%%%%%%%%%%%%%%%%%%%%

Pycnonuclear fusion reactions are nuclear fusion reactions between heavy atomic nuclei that occur in regions of high density and are schematically expressed as $_{A}^{Z}\textrm{X} + \, _{A}^{Z}\textrm{X} \to \, _{2A}^{2Z}\textrm{Y}$. Examples of this type of reaction are the fusion of carbon with carbon ($^{12}\textrm{C} \, + \, ^{12}\textrm{C}$) and oxygen with oxygen ($^{16}\textrm{O} \, + \, ^{16}\textrm{O}$). These fusion reactions are significantly density sensitive but nearly temperature independent \citep{2005PhRvC..72b5806G}. At these high densities, the probability of nuclei undergoing the tunneling process becomes considerable, and therefore fusion reactions are likely to occur often enough to begin to affect the star's stability.

It is well known that pycnonuclear fusion reactions make WDs unstable. However, the nuclear fusion rates at which these reactions occur are highly uncertain  because of some poorly constrained parameters \citep[see, e.g.,][]{2005PhRvC..72b5806G, 2006PhRvC..74c5803Y}. However, the occurrence of these reactions has already been theoretically analyzed for a considerable density range that includes values of densities that are found inside WDs \citep{2013PhRvD..88h1301C,2017MNRAS.469...95C}. \cite{2005PhRvC..72b5806G}, for example, developed a set of analytical equations to calculate the rates of pycnonuclear reactions between identical atomic nuclei and applied them to carbon fusion inside WDs. The authors found that carbon burning becomes important for a temperature $T \sim (4-15) \times 10^{8}$~K assuming a density $\rho \lesssim 3 \times 10^{9}$~g~cm$^{-3}$, and for $\rho \sim (3-50) \times 10^{9}$~g~cm$^{-3}$ considering $T \lesssim 10^ {8}$~K.

Several studies have been carried out with the aim of performing a stability analysis of the matter in the cores of WDs against pycnonuclear fusion reactions and electron capture reactions \citep[see, e.g.,][]{2013ApJ...762..117B, otoniel_strongly_2019, 2021ApJ...921..138N}. In a recent work, \cite{Otoniel2021} investigated the stability of the rapidly rotating WD of the CTCV J2056–3014 system, where they determined mass density limits for which pycnonuclear fusion reactions start, thus inferring a maximum value for the mass of this WD. The authors also studied the impact of rapid rotation on the structure and stability of the WD to calculate a minimum mass value for this remnant. The procedure performed in this work to find the mass limits of the WDs follows the methodology presented in that study.

Therefore, to obtain the upper mass limit, we assume that WD has a uniform chemical composition throughout the star, and we investigate the stability of the WD matter to pycnonuclear fusion reactions using up-to-date theoretical models \citep[see, e.g.,][]{2005PhRvC..72b5806G,2009PhRvC..80a5804G}. These models analyze nuclear fusion rates in dense matter, covering various burning regimes. By addressing the Coulomb barrier problem and proposing a versatile formula for reaction rates, they offer a comprehensive solution adaptable to current theoretical uncertainties. The model's efficacy is demonstrated through the examination of the $^{12}\text{C} + \, ^{12}\text{C}$ fusion reaction, crucial for understanding stellar phenomena like nuclear burning in evolved stars, type Ia supernovas, and accreting neutron stars. Leveraging a parameter-free model for the energy-dependent S factor calculation, accounting for Pauli nonlocality effects, enhances accuracy in estimating fusion efficiencies, particularly emphasizing carbon ignition at densities exceeding $10^9$~g~cm$^{-3}$ \citep{2005PhRvC..72b5806G,2009PhRvC..80a5804G}.

%%%%%%%%%%%%%%%%%%%%%%%%%%%%%%%%%%%%%%%%%%%%%%%%%%%%%%%%%%%%%%%%%%%%%%%
%%%%%%%%%%%%%%%%%%%%%%%%%%%%%%%%%%%%%%%%%%%%%%%%%%%%%%%%%%%%%%%%%%%%%%%
\section{Gravitational emission mechanism}
\label{sec:GW_emission}
%%%%%%%%%%%%%%%%%%%%%%%%%%%%%%%%%%%%%%%%%%%%%%%%%%%%%%%%%%%%%%%%%%%%%%%
%%%%%%%%%%%%%%%%%%%%%%%%%%%%%%%%%%%%%%%%%%%%%%%%%%%%%%%%%%%%%%%%%%%%%%%

We investigated the gravitational radiation emitted by fast-spinning magnetic WDs with structural deformation caused by matter accretion. Thus, within this GW generation mechanism, we consider a WD accreting matter through magnetic poles that do not coincide with the spin axis of the star. In this scenario, the WD receives matter via an accretion column, which may come from a remnant debris disk generated after the merger of two WDs \cite[see e.g.,][]{2014MNRAS.438...14D,2012ApJ...749...25G} or from a secondary star that transfers part of its mass through a disk truncated by field lines, or even without forming a disk \cite[see e.g.,][]{1998MNRAS.298..285W,2003cvs..book.....W}. In this way, the WD accumulates an amount of matter at its magnetic poles causing irregularities in the shape of the star.

Below, we describe the mechanism of matter accretion and deduce the GW amplitude emitted in this process. Moreover, we found an expression for amplitude that takes into account the effect of magnetic deformation on GW emission.

%%%%%%%%%%%%%%%%%%%%%%%%%%%%%%%%%%%%%%%%%%%%%%%%%%%%%%%%%%%%%%%%%%%%%%%
\subsection{Accretion of matter: basic equations}
\label{sec:acc_mech} % used for referring to this section from elsewhere
%%%%%%%%%%%%%%%%%%%%%%%%%%%%%%%%%%%%%%%%%%%%%%%%%%%%%%%%%%%%%%%%%%%%%%%

We consider a rotating rigid object whose axes of symmetry ($x_{1}$, $x_{2}$, $x_{3}$) rotate with the object, and the corresponding principal moments of inertia are $I_{1}$ , $I_{2}$ and $I_{3}$, respectively. This solid rotates with angular velocity $\Omega$ relative to the axis $x'_{3}$ of an inertial frame whose axes are ($x'_{1}$, $x'_{2}$, $x'_ {3}$). We also consider that the axis of the magnetic dipole is given by the axis $x_3$ and that this axis forms an angle $\alpha$ with the axis $x'_3$. In other words, the angle of inclination of the magnetic axis relative to the axis of rotation is given by $\alpha$. Thus, making $I_{1} = I_{2}$ and $d$ being the distance to the emitting source, the GW amplitude, $h_{ac}$, can be represented by \citep{shapiro/2008, maggiore/2008}

\begin{equation}
  h_{ac} = \frac{4G}{c^{4}} \frac{(I_{1}-I_{3}) \Omega ^{2}}{d} \sin ^{2}\alpha.
  \label{AmpAC2}
\end{equation}

It is worth mentioning that the amplitude equation above is deduced considering that the angle between the axis of rotation ($x'_{3}$) and the line of sight is zero. With this assumption, we maximize the value of GW amplitude when we consider only this parameter \cite[see][for details on this dependency]{maggiore/2008}.

Now, we need to determine the moments of inertia $I_{1}$ and $I_{3}$. Assuming a deformation created only by accretion of mass, we can consider that the star has deformities or mountains of matter on the $x_{3}$ axis, so that these deformities can be approximated to small spheres of masses close to the star's magnetic poles. Thus, with this configuration, the term $I_{1}-I_{3}$ and the GW amplitude can be expressed as \citep{2020MNRAS.492.5949S}

\begin{equation}
  I_{1}-I_{3} = 2\delta m ~R^{2},
    \label{I1I3ACR}
\end{equation}

\begin{equation}
  h_{ac} = \frac{8G}{c^{4}} \frac{\delta m ~R^{2} \Omega ^{2}}{d} \sin ^{2}\alpha,
\label{AmpAC4}
\end{equation}
where $R$ is the radius of the star and $\delta m$ is the amount of mass accumulated on one magnetic pole.

However, together with the deformations caused by $\delta m$, we can also include the effect of the magnetic field on the spherical shape of the star. In other words, we can find an equation that expresses the irregularities in the star's structure due to accretion and the intense magnetic field. For this task, we must assume that the star, previously considered as a sphere in the above assumption, contracts along the magnetic field, becoming an oblate spheroidal shape. Thus, in addition to the small spheres we assume in the above configuration, we must consider the star as an oblate object.

Assuming that $R$ is the polar radius of the oblate spheroid ($x_{3}$ axis) and $R_{\rm eq} = R + b$ is the radius at the star's equator ($x_{1}$ and $x_{2}$ axes), we can write the new moments of inertia $I_{1}$ and $I_{3}$ as

\begin{equation}
  I_{1} = \frac{1}{5} M(R_{\rm eq}^{2} + R^{2}) + 2\,\delta m\, R^{2},
\label{I1ACnew}
\end{equation}

\begin{equation}
  I_{3} = \frac{2}{5} MR_{\rm eq}^{2} + 2 \frac{2}{5} \delta m\, a^{2} \approx \frac{2}{5} MR_{\rm eq}^{2},
\label{I3ACnew}
\end{equation} 
where $M$ is the mass of the star and $a$ is the radius of the amount of mass accumulated on one magnetic pole. In the last term of equation (\ref{I3ACnew}), we assume $R_{\rm eq} \gg a$. The first term on the right side of equations  (\ref{I1ACnew}) and  (\ref{I3ACnew}) corresponds to the moment of inertia of an oblate spheroid, whereas the second term represents the moment of inertia of the two mountains of matter on the $x_{3}$ axis. 

Now, considering that $R_{\rm eq} \gg b$, the term $I_{1}-I_{3}$ can be expressed as follows

\begin{equation}
    I_{1}-I_{3} = 2\,\delta m \, R^{2} - \frac{2}{5} M\,R^{2} \frac{b}{R}.
\label{I1I3new1}
\end{equation} 

By recalling that the moment of inertia for a sphere of radius $R$ is given by $I = (2/5) MR^2$ and that the definition of the eccentricity $\varepsilon = \sqrt{R_{\rm eq}^ 2 - R^2}/R$ returns $\varepsilon = b/R$, one obtains

\begin{equation}
    I_{1}-I_{3} = 2\,\delta m \, R^{2} - I\varepsilon.
\label{I1I3new2}
\end{equation}

It is worth noting that the first term on the right side of the equation~(\ref{I1I3new2}) indicates the deformation produced by the accretion of matter mechanism, while the second term on the right side represents the deformability due to the intense magnetic field. Furthermore, the minus sign between these two terms should be noted, suggesting that the two GW emission effects act to cancel each other out. That is, the eccentricity created by the magnetic field does not contribute to the accretion process to produce GWs with larger amplitudes but rather weakens the wave amplitude. This occurs because the deformations produced by the two processes appear on different semi-axes, reducing the irregularities in the star shape.

Substituting equation (\ref{I1I3new2}) into equation (\ref{AmpAC2}), it follows that

\begin{equation}
  h_{\rm acmg} = \frac{4G}{c^{4}} \frac{ \left| 2\,\delta m \, R^{2} - I\varepsilon \right| \Omega ^{2}}{d} \sin ^{2}\alpha,
  \label{AmpACDF}
\end{equation}
where the modulus function appears because $h_{\rm acmg}$ is defined as the strain and we introduce the $acmg$ index to represent the amplitude that takes into account the effect of the magnetic field. Moreover, the eccentricity caused by the magnetic field can also be represented as \citep[see, e.g.,][]{1996A&A...312..675B,2000A&A...356..234K,2006A&A...447....1R}

\begin{equation}
\varepsilon = \kappa \frac{B^{2} R^{4}}{G M^{2}},
    \label{excentridade1}
\end{equation}
where, as before, $B$ is the magnetic field strength on the star’s surface and $\kappa$ is the distortion parameter, which depends on the magnetic field configuration and equation of state (EoS) of the star.

Thereby, we find an equation that represents the GW emission for the process of accretion of matter together with the mechanism of magnetic deformation that depend on the amount of mass accumulated on the WD's magnetic poles and also on the eccentricity caused by the magnetic field.

%%%%%%%%%%%%%%%%%%%%%%%%%%%%%%%%%%%%%%%%%%%%%%%%%%%%%%%%%%%%%%%%%%%%%%%
%%%%%%%%%%%%%%%%%%%%%%%%%%%%%%%%%%%%%%%%%%%%%%%%%%%%%%%%%%%%%%%%%%%%%%%
\section{Results and Discussions}
\label{sec:res_dis}
%%%%%%%%%%%%%%%%%%%%%%%%%%%%%%%%%%%%%%%%%%%%%%%%%%%%%%%%%%%%%%%%%%%%%%%
%%%%%%%%%%%%%%%%%%%%%%%%%%%%%%%%%%%%%%%%%%%%%%%%%%%%%%%%%%%%%%%%%%%%%%%

Here we extend our previous study \citep{2020MNRAS.492.5949S} by expanding the number of fast-spinning magnetic WDs and establishing constraints on the mass and radius for these WDs from the construction of stability regions. These limits are determined by the Kepler frequency and by the pycnonuclear fusion reactions. We also approach the effect of the deformation of the WD produced by the intense magnetic field on the GW amplitude emitted by the accretion mechanism.

Table \ref{tab:Tab_Param} shows the six fast-spinning magnetic WDs that we analyzed in this work along with their main parameters. These sources are five binary systems: LAMOST J0240+1952, CTCV J2056-3014, AE Aqr, V1460 Her, RX J06480-4418; and an AXP: 4U 0142+61 for which we assume the model described by \citet{Borges2020} in which this object is a massive remnant WD accreting from a debris disk formed during the merger of two other WDs. The magnetic field strength of this AXP is in an interval inferred from the rotation variation due to the coupling between the magnetic field and the disk to reproduce the spindown value of this object \citep{Borges2020}.

\begin{table*}
\centering
\caption{Six fast-spinning magnetic WDs along with their main parameters: Period (P), magnetic field (B), Temperature (T) and distance to source (d).}
\label{tab:Tab_Param}
\begin{tabular}{|lcccc|}
\hline
\textbf{SYSTEMS} & \textbf{\begin{tabular}[c]{@{}c@{}}P\\(s)\end{tabular}} & \textbf{\begin{tabular}[c]{@{}c@{}}B \\(G)\end{tabular}} & \textbf{\begin{tabular}[c]{@{}c@{}}T \\(K)\end{tabular}} & \textbf{\begin{tabular}[c]{@{}c@{}}d \\(pc)\end{tabular}} \\ \hline
AXP 4U 0142+61 $^{\rm a,b}$ & 8.69 & $5.0 \times 10^{7}$ $^{\rm b}$ & $10^{5}$ & 3780 $^{\rm a}$ \\
RX J06480-4418 $^{\rm c}$ & 13.18 & $10^{8}$ & $10^{5}$ & 520.94 $^{\rm h}$ \\
LAMOST J0240+1952 $^{\rm d}$ & 24.93 & $10^{7}$ & $\leqslant 2.5 \times 10^{4}$ & 592.98 $^{\rm h}$ \\
CTCV J2056-3014 $^{\rm e}$ & 29.61 & $10^{7}$ & $10^{4}$ & 256.29 $^{\rm h}$ \\
AE Aquarii $^{\rm f}$ & 33.08 & $5.0 \times 10^{7}$ & $10^{7}$ & 91.87 $^{\rm h}$ \\
V1460 Herculis $^{\rm g}$ & 38.87 & $10^{7}$ & $10^{3}$ & 266.03 $^{\rm h}$ \\ \hline
\end{tabular}

$^{\rm a}$ \citet{Olausen&Kaspi2014}; $^{\rm b}$ \citet{Borges2020}; $^{\rm c}$ \citet{Israel1997,Mereghetti2009,Mereghetti2021,Mereghetti2011}; $^{\rm d}$ \citet{Pelisoli2022}; $^{\rm e}$ \citet{LopesDeOliveira2020,Otoniel2021}; $^{\rm f}$ \citet{Terada2008,Kitaguchi2014}; $^{\rm g}$ \citet{Ashley2020,KJURKCHIEVA2017}; $^{\rm h}$ \citet{Gaia2022}

\end{table*}

Furthermore, to analyze the effect of the magnetic field on the GW amplitude emitted by the accretion mechanism, we established a value of $B = 10^{7}$~G for the WDs that do not have known magnetic field values, such as the LAMOST J0240+1952 and V1460 Her. This is a moderate field value for magnetic cataclysmic variables \cite[see e.g., Figure 30 by][]{2020AdSpR..66.1025F} and one that is also not so intense that the magnetic interaction with the secondary star locks the WD into a rotation synchronized with the orbital period. For CTCV J2056-3014, although magnetic field estimates show that this WD has possibly a low field value ($\sim 10^{4}$~G; \cite{LopesDeOliveira2020}), we also assume here a magnetic field of $B = 10^{7}$~G, in order to investigate the influence of an intense magnetic field on the emission of gravitational radiation from this system.

Using the approach described in Sec. \ref{sec:model}, we numerically calculate the minimum mass (maximum radius) and maximum mass (minimum radius) from the WD stability conditions. This approach is presented in \citet{Otoniel2021}, where the authors model a WD rotating uniformly within Hartle's formalism for stellar rotation, considering an EoS that takes into account not only the contribution of a Fermi gas but also the contributions of electron-ion, electron-electron and ion-ion interactions, and explore instabilities due to the effects of rapid rotation and microscopic stability on the structure of WDs.

Table \ref{tab:Tab_Mass} presents the mass-radius limits found for each WD of the systems presented in Table \ref{tab:Tab_Param}. The minimum mass values are obtained assuming the WD is rotating near its mass-shedding limit, while the maximum mass values are calculated from the occurrence of pycnonuclear reactions. The WD composition adopted in each simulation is also presented in this table. It is worth mentioning that the composition of WDs is not well established. Therefore, we simulate canonical rotating WDs composed purely of C, except for those that show some indication of a different chemical composition or having already been modeled as WDs composed of more than one chemical element. We simulate, for example, the systems RX~J06480-4418 and 4U~0142+61 twice to cover two different compositions. We did this to investigate whether the mass-radius limits can vary significantly with this parameter.

\begin{table*}
\centering
\caption{The minimum (maximum radius) and maximum (minimum radius) mass values of some fast-spinning magnetic WDs inferred from the stability conditions. The minimum mass value is obtained assuming that the WD rotation must be close to the mass-shedding limit, while the maximum mass value is calculated from the occurrence of pycnonuclear reactions. The WD composition in each case is also shown.}
\label{tab:Tab_Mass}
\begin{tabular}{|lccccc|}
\hline
\textbf{SYSTEMS} & \textbf{\begin{tabular}[c]{@{}c@{}}M$_{\rm min}$\\ ($M_{\odot}$)\end{tabular}} & \textbf{\begin{tabular}[c]{@{}c@{}}R$_{\rm max}$\\ ($10^8$ cm)\end{tabular}} & \textbf{\begin{tabular}[c]{@{}c@{}}M$_{\rm max}$\\ ($M_{\odot}$)\end{tabular}} & \textbf{\begin{tabular}[c]{@{}c@{}}R$_{\rm min}$\\ ($10^8$ cm)\end{tabular}} & \textbf{Composition} \\ \hline
LAMOST J0240+1952 & 0.653 & 10.067 & 1.380 & 1.307 & C \\
CTCV J2056-3014 & 0.554 & 10.867 & 1.380 & 1.309 & C \\
AE Aquarii & 0.545 & 11.214 & 1.380 & 1.307 & { } C $^{\rm a}$ \\
V1460 Herculis & 0.433 & 11.775 & 1.379 & 1.318 & C \\
AXP 4U 0142+61 & 1.136 & 6.325 & 1.380 & 1.312 & C  \\
AXP 4U 0142+61 & 1.132 & 6.310 & 1.377 & 1.312 & C 50\%/O 50\%  $^{\rm b}$ \\
RX J06480-4418 & 0.938 & 7.703 & 1.376 & 1.311 & C 50\%/O 50\% $^{\rm c}$ \\
RX J06480-4418 & 0.932 & 7.686 & 1.369 & 1.308 & O 50\%/Ne 50\% $^{\rm c}$ \\ \hline
\end{tabular}

$^{\rm a}$ \cite{Kitaguchi2014}; $^{\rm b}$ \cite{Borges2020}; $^{\rm c}$ \cite{Mereghetti2021,Mereghetti2011}

\end{table*}

It can be seen from Table~\ref{tab:Tab_Mass} that the minimum and maximum mass values, as well as the maximum and minimum radius values, of AXP~4U~0142+61 and RX~J06480-4418 change very little for the different compositions of the WD. These values differ only to the second or third decimal place. Thus, when we consider the GW amplitude, this difference in mass and radius is not relevant, showing that the amplitude value is practically the same for both cases. Thus, we chose only one of the values for these sources in the analysis of gravitational radiation. We then select AXP~4U~0142+61 with a composition of C and RX~J06480-4418 with a composition of C(50$\%$)/O(50$\%$).

Now, we can proceed with the estimate of the GW emission considering the parameters of the systems presented in Tables \ref{tab:Tab_Param} and \ref{tab:Tab_Mass}. We use equation~(\ref{AmpACDF}) to calculate the GW amplitude from the effects of accretion and magnetic field strength on the star's structure. This equation shows that the GW amplitude depends on the amount of accumulated mass; however, it is not easy to predict the amount of matter that may have been accreted by WD and how much of it was dispersed on its surface. In non-magnetic systems, the accreted material on the surface of the WD can reach $10^{-4} \, \rm{M}_{\odot } - 10^{-6} \, \rm{M}_{\odot } $; nonetheless, the amount of accumulated mass can be reduced if accretion onto the WD is confined to a smaller fraction of its surface area \citep[see, e.g.,][]{1980IAUS...88..431M, 1982ApJ...261..649S, 2020A&A...634A...5J, 2022Natur.604..447S}. Here we attribute four values to the amount of matter deposited at the magnetic poles: $\delta m =$ ($10^{-3}\, \rm{M}_{\odot } $, $10^{-4} \, \rm{M}_{\odot } $, $10^{-5}\, \rm{M}_{\odot } $, $10^{-6}\, \rm{M}_{\odot } $). These values are assessed to determine the necessary accreted mass for generating gravitational radiation detectable by space-based GW instruments. It is observed that lesser accumulations of matter may result in GW amplitudes that could potentially escape detection, as we will see below. Furthermore, we consider a range for the angle of inclination $\alpha$ from $10^{\circ}$ to $90^{\circ}$.

However, before evaluating the GW amplitude values, we must understand the effect of the deformation created by the magnetic field on the amplitude generated by the mountain of matter due to accretion. In Sec.~\ref{sec:acc_mech}, we note that the irregularity formed by the magnetic field acts by decreasing the wave amplitude produced by accretion (see equations (\ref{I1I3new2}) and (\ref{AmpACDF})). Therefore, if the deformation effects created by the two processes become similar, they may cancel each other and emit no GWs. 

Thus, to determine the values of the parameters for which the equation (\ref{I1I3new2}) is null, an expression that represents the dependence of $\delta m$ in relation to $B$, $M$ and $R$ can be found, namely

\begin{align}
    2\,\delta m \, & R^{2} - I\varepsilon = 0, \nonumber \\
    \delta m &= \frac{1}{5} \, \kappa  \, \frac{B^{2} R^{4}}{G M}.
\label{deltam}
\end{align}

\begin{figure}

	\includegraphics[width=\columnwidth]{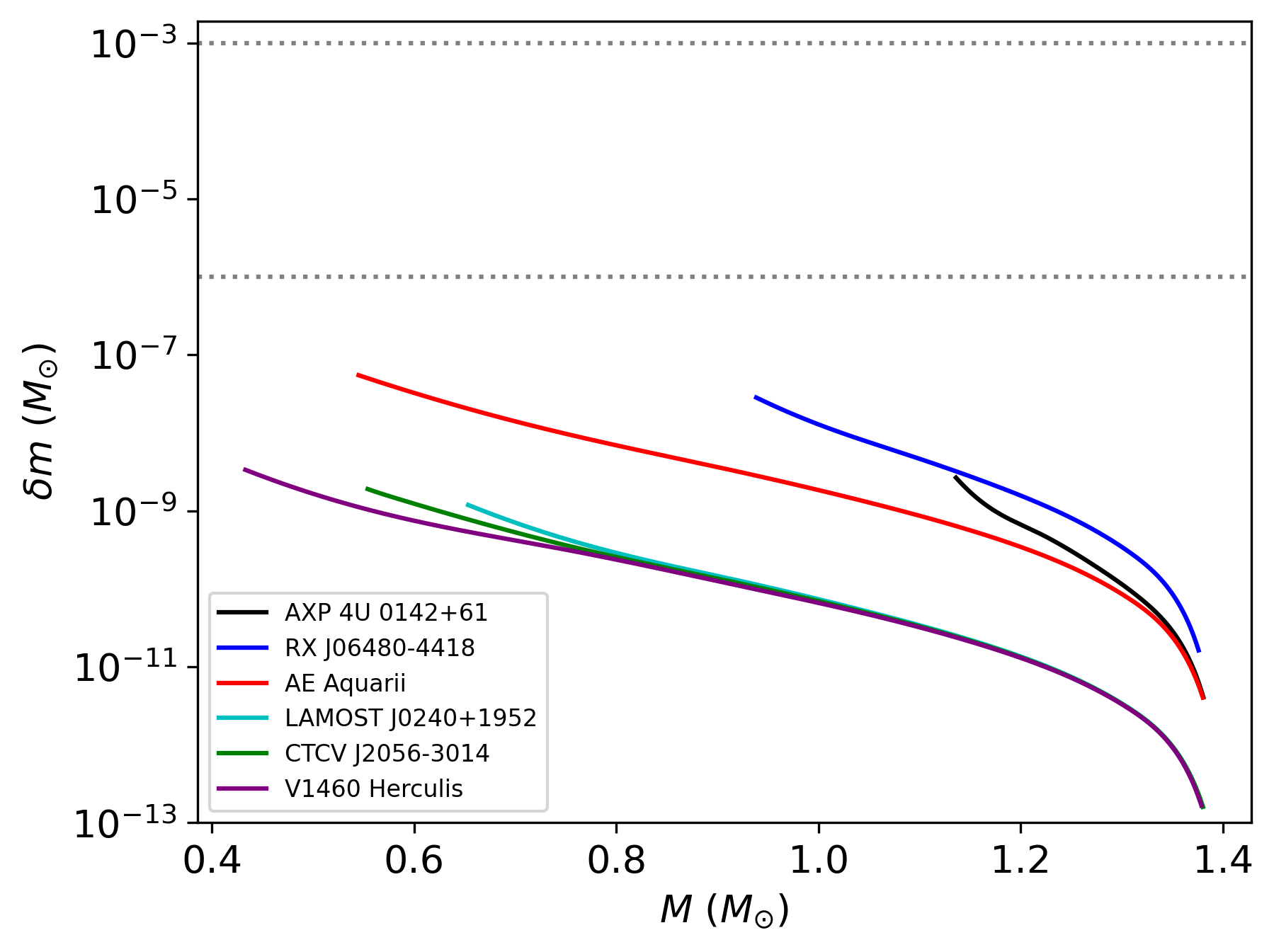}

    \caption{Relation between $\delta m$ and $M$, given by equation~(\ref{deltam}), for which $h_{\rm acmg} = 0$, that is, when the deformation in the star structure created by accretion and magnetic field are equal. The mass range presented by the solid curves is established by the $M_{\rm min}$ and $M_{\rm max}$ of each system. The dotted lines form a $\delta m$ range that contains the four considered $\delta m$ values.}
    \label{fig:DeltamM}
\end{figure}

Here, we assign the value of $B$ from Table \ref{tab:Tab_Param} and we adopt $\kappa\simeq 10$, which is a conservative value that holds for an incompressible fluid star with a dipole magnetic field \citep[see, e.g.][]{1954ApJ...119..407F}. Similar values are obtained when relativistic models based on a polytropic EoS with dipole magnetic field are considered \citep[see, e.g.][]{2000A&A...356..234K}. Thereby, the equation for $\delta m$ is only a function of the mass/radius and we can calculate the value of $\delta m$ for which $h_{\rm acmg} = 0$ using the established mass interval by the $M_{\rm min}$ and $M_{\rm max}$ of each source. The result of this analysis is shown in Fig. \ref{fig:DeltamM} where the dotted lines form the interval of $\delta m$ that contains the four values of $\delta m$ that we considered above.

It can be observed from this figure that all the analyzed systems are below the interval of $10^{-6} M_{\odot } \leqslant \delta m \leqslant 10^{-3} M_{\odot }$ and, thus, the GW amplitude is not zero for any value in the mass range of these WDs when we take these $\delta m$ values into account. Therefore, the effect of accretion on the WD shape, and consequently on the GW amplitude, is dominant when compared to magnetic effects on the structure of the star. For $h_{\rm acmg} = 0$ at some point in the mass range, these WDs must have $\delta m \lesssim 10^{-7} M_{\odot }$ for RX J06480-4418 and AE Aqr, and $\delta m \lesssim 10^{-9} M_{\odot }$ for LAMOST J0240+1952, CTCV J2056-3014, V1460 Her and 4U 0142+61. With this, we realize that the GW amplitude of these sources is not significantly affected by the deformation produced by the magnetic field. It is worth mentioning that the mass accumulation at the magnetic poles is significantly influenced by the interplay between the mass-accretion rate and other factors, such as the magnetic field \citep[see, e.g.,][]{2022MNRAS.514L..11S}. This interaction can lead to smaller values of $\delta m$ than those mentioned previously, sometimes resulting in the cancellation of the effects of deformations caused by accretion and the magnetic field, or with the magnetic field exerting a greater effect. Nevertheless, the resulting GW amplitudes in both cases do not generate observable signals for spatial detectors due to the small $\delta m$ values and the insufficient ellipticity generated by the analyzed sources' magnetic field intensity \citep[see][]{2020MNRAS.492.5949S}.

\begin{figure*}
	\includegraphics[width=\columnwidth]{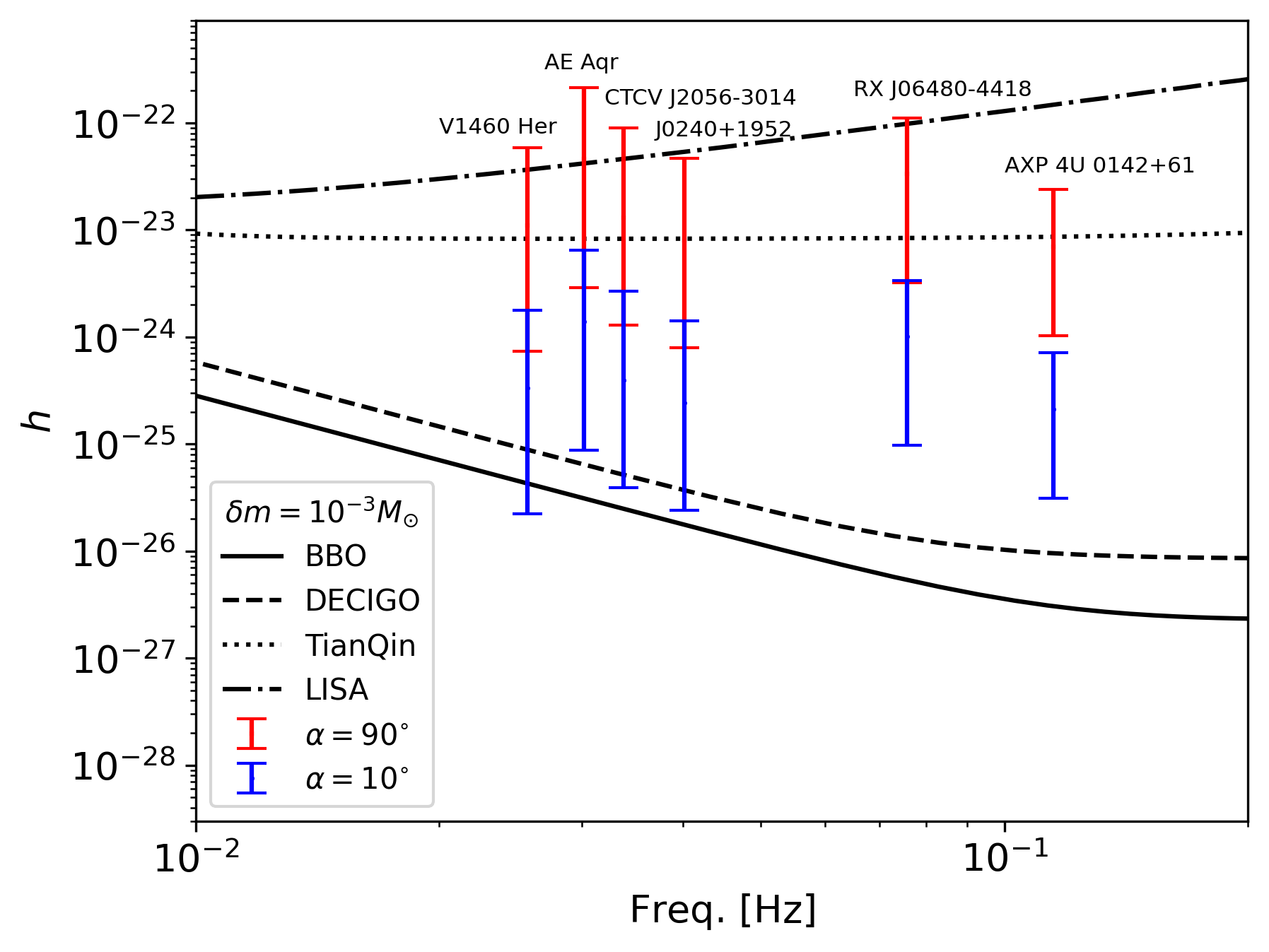}
    \includegraphics[width=\columnwidth]{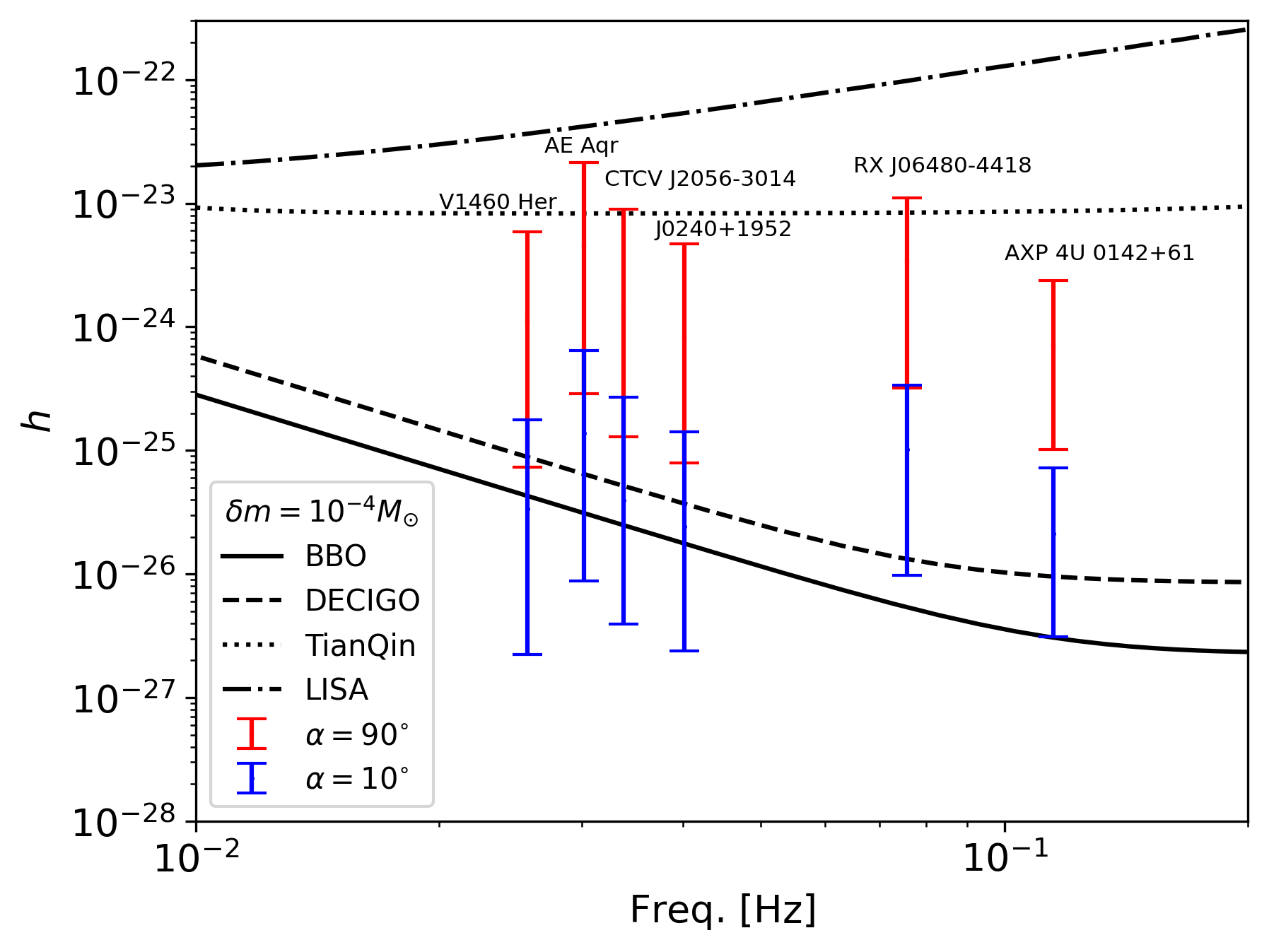}
    %\vspace{0.9cm}
    \includegraphics[width=\columnwidth]{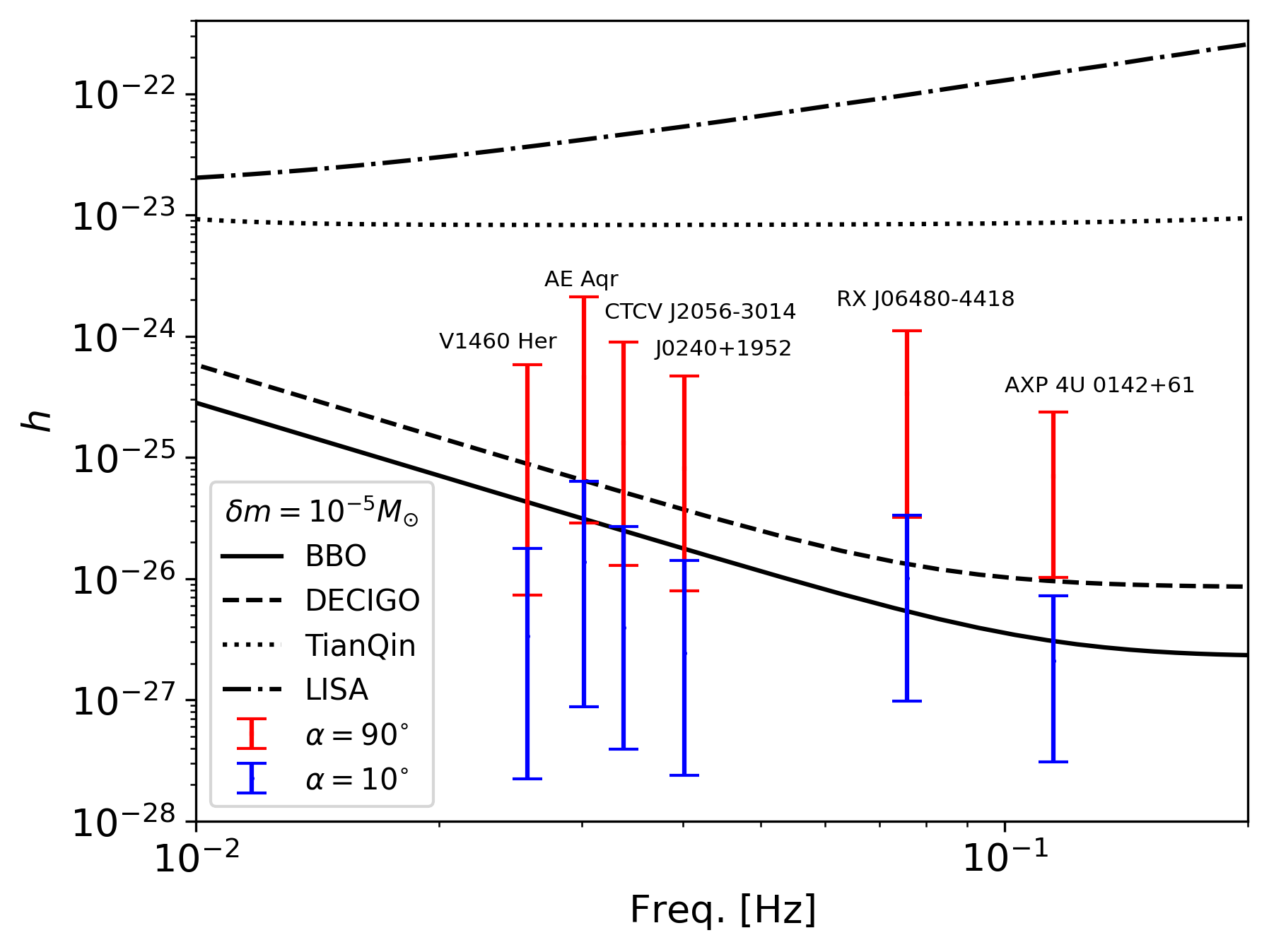}
    \includegraphics[width=\columnwidth]{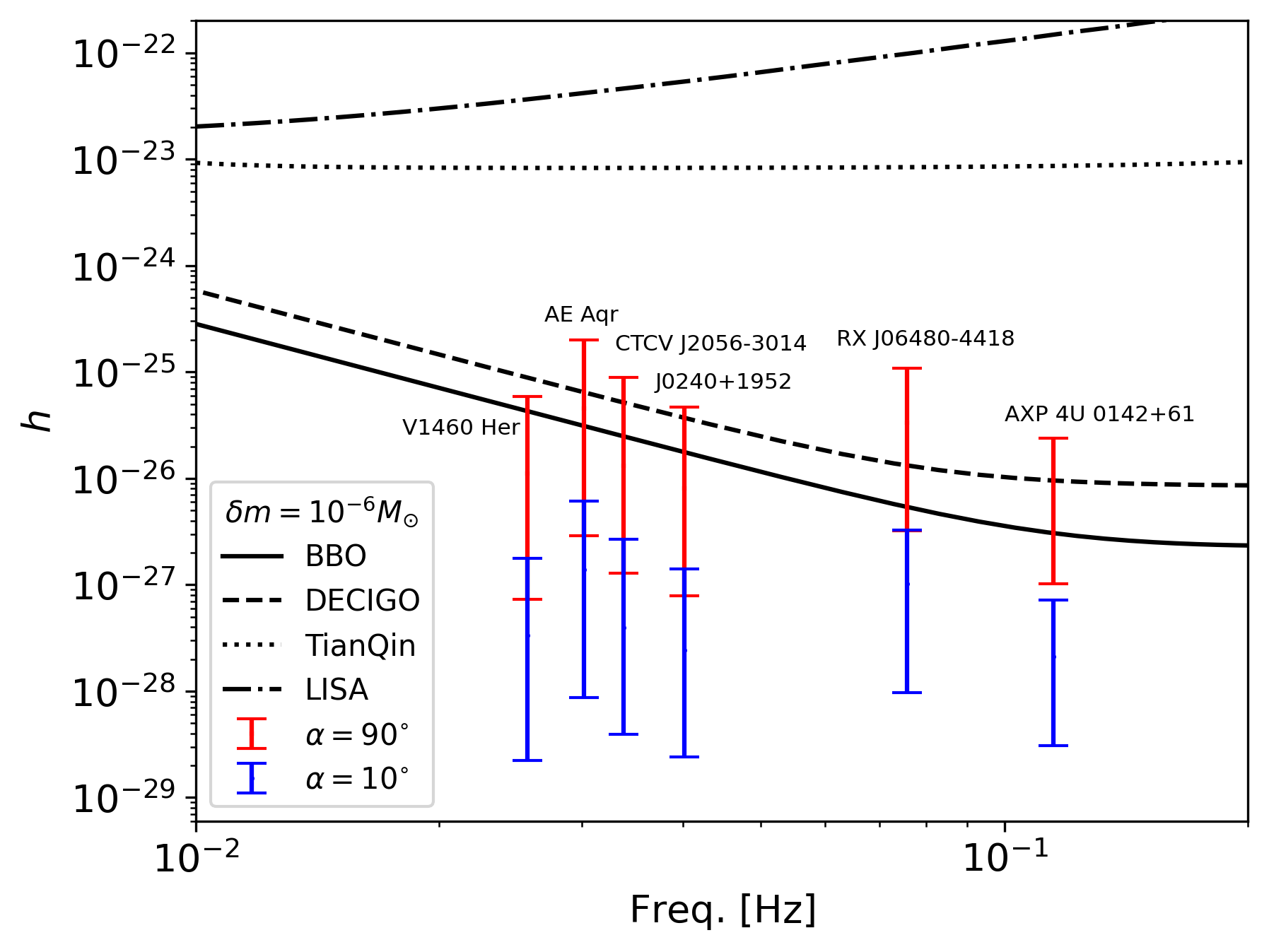}
    \caption{GW amplitude for different mass values taking into account $\alpha = 90^{\circ}$ (red color)  and $\alpha = 10^{\circ}$ (blue color). The vertical bars stands for  M$_{\rm min} \leq$ M$_{\rm WD} \leq$ M$_{\rm max} $, from top to bottom. The different panels represent a quantity of accreted mass at the WD's magnetic poles, $\delta m =$ ($10^{-3}, 10^{-4}, 10^{-5}, 10^{-6}$) $ M_{\odot }$. The sensitivity curves of the space-based detectors LISA, TianQin, BBO and DECIGO are shown considering an SNR $= 8$ and $\tau_{\rm obs} = 1$~yr.}
    \label{fig:AmpVSFreq}
\end{figure*}

With the issue discussed above, we can now proceed to calculate the GW amplitude values for the six fast-spinning magnetic WDs considered in this work. Using equation~(\ref{AmpACDF}) and assuming the mass range established by $M_{\rm min}$ and $M_{\rm max}$ from Table~\ref{tab:Tab_Mass}, the GW amplitude is presented in Fig.~\ref{fig:AmpVSFreq} for each system, where we set two values for the inclination angle, $\alpha = 90^{\circ}$ and $\alpha = 10^{\circ}$. We chose these two values of $\alpha$ in order to analyze the amplitude values for a small inclination ($10^{\circ}$), close to the alignment of the axes of rotation and magnetic field, and for an angle ($90 ^{\circ}$) at which the GW amplitude value is greatest. Figure~\ref{fig:AmpVSFreq} displays four panels where the GW amplitude is presented as a function of the frequency of the WDs, so that the vertical bars represent $M_{\rm min} \leq M \leq M_{\rm max}$ , from top to bottom. In each panel, a different value of $\delta m$ is considered. We also show the sensitivity curves of the space-based detectors LISA, TianQin, BBO and DECIGO \citep{2019CQGra..36j5011R, 2019PhRvD.100d4042L, 2021PTEP.2021eA107M, 2017PhRvD..95j9901Y}. It is worth mentioning that to plot the sensitivity curves, we use the minimum amplitude, $h_{min}$, that can be measured by the detector, for a periodic signal, for a given signal-to-noise ratio (SNR) and observation time $\tau_{\rm obs}$ 
\citep[see section 7.6 and equation (7.134) of][]{maggiore/2008}. Thereby, the sensitivity curves are set to SNR~$= 8$ and $\tau_{\rm obs} = 1$ year.

Observing Fig.~\ref{fig:AmpVSFreq}, we notice that LISA possibly does not have enough sensitivity to observe the fast-spinning magnetic WDs analyzed in this study. AE Aqr, for example, which is a standout font for this detector, would need to have $\delta m \approx 10^{-3}\, \rm{M}_{\odot}$, an angle of inclination $ \sim 90^{\circ}$ and mass values close to $M_{\rm min}$ to be detected by this instrument. A difficult combination of parameters since $\delta m \approx 10^{-3}\, \rm{M}_{\odot}$ is already big enough for a mountain of matter accreted in a WD. TianQin, on the other hand, would observe all fonts if they had $\delta m \approx 10^{-3}\, \rm{M}_{\odot}$, $\alpha \sim 90^{\circ} $ and were not so massive; or it would only detect AE Aqr and RX J06480-4418 systems with $\delta m \approx 10^{-4}\, \rm{M}_{\odot}$, $\alpha \sim 90^{\circ} $ and with mass close to $M_{\rm min}$.

For BBO and DECIGO, the results are more optimistic. Assuming $\alpha \sim 90^{\circ}$, for example, these instruments would observe WDs over the entire mass range as long as $\delta m \geq 10^{-4} \, \rm{M}_ {\odot}$. For a $\delta m \approx 10^{-5}\, \rm{M}_{\odot}$, only the RX~J06480-4418 and 4U~0142+61 systems are detectable for the entire mass range. For a $\delta m \approx 10^{-6}\, \rm{M}_{\odot}$, systems must have small mass values to be observed, except for RX~J06480-4418 and 4U ~0142+61 which can have larger mass values. When we assume $\alpha \sim 10^{\circ}$, the WDs will hardly be detected by BBO and DECIGO if they have $\delta m \leq 10^{-5}\, \rm{M}_{\odot} $, except RX~J06480-441 and 4U~0142+61, which are measurable for some mass values $M$ considering $\delta m \approx 10^{-5}\, \rm{M}_{\odot }$.

In general, we noticed that regarding the emission of GWs from matter accretion mechanism, the six fast-spinning and magnetic WDs are unlikely to be observed by the LISA and TianQin detectors and, thus, are not candidate sources for these instruments. However, these systems are good candidates for BBO and DECIGO and may be detected if they have a $\delta m \geq 10^{-5}~\rm{M}_{\odot}$. For an amount of $\delta m$ smaller than this value, only the 4U~0142+61 and RX~J06480-441 systems stand out, since for $\delta m = 10^{-6}~\rm{M }_{\odot}$, these systems are still detected if we assume an inclination angle $\alpha$ close to the maximum value.

The results obtained here for the AE~Aqr and RX~J06480-4418 systems are in agreement with the results presented in \cite{2020MNRAS.492.5949S}. However, in this study, it was possible to obtain an amplitude interval for a value of $\delta m$, in which we can observe the dependence of this parameter with the mass/radius of the WD, with the inclination of the magnetic axis and with the strength of the magnetic field.

%%%%%%%%%%%%%%%%%%%%%%%%%%%%%%%%%%%%%%%%%%%%%%%%%%%%%%%%%%%%%%%%%%%%%%%
%%%%%%%%%%%%%%%%%%%%%%%%%%%%%%%%%%%%%%%%%%%%%%%%%%%%%%%%%%%%%%%%%%%%%%%
\section{Conclusions}
\label{sec:concl}
%%%%%%%%%%%%%%%%%%%%%%%%%%%%%%%%%%%%%%%%%%%%%%%%%%%%%%%%%%%%%%%%%%%%%%%
%%%%%%%%%%%%%%%%%%%%%%%%%%%%%%%%%%%%%%%%%%%%%%%%%%%%%%%%%%%%%%%%%%%%%%%

In this article, we have investigated the emission of GWs from a class of stellar remnants that has attracted great attention from the astrophysics community in recent years: the fast, massive, and magnetic WDs. These stars, which can be found both isolated and in binary systems, have a high magnetic field ranging from $10^6$~G to $10^9$~G and a short rotation period of around seconds to a few minutes. Some of these WDs are also classified as WD pulsars because of their pulsed electromagnetic emission, similar to that found in NS pulsars. Therefore,  we analyze the continuous gravitational radiation from these sources using two mechanisms: matter accretion and magnetic deformation. In both mechanisms, the GW emission is generated by the asymmetry around the axis of rotation of the star due to the accumulated mass at the magnetic poles and the intense magnetic field, respectively.

Our efforts in this scope started with \cite{2020MNRAS.492.5949S}, where we investigated the continuous gravitational radiation from three fast-spinning magnetized WDs; however, here we extend our analysis by applying it to a larger number of sources, establishing constraints on the WD mass and radius from the construction of stability regions, and modifying some system parameters such as the inclination of the magnetic field in relation to the axis of rotation. For this, we investigated the GW emission in six fast and magnetic WDs: five in binary systems (LAMOST~J0240+1952, CTCV~J2056-3014, AE~Aqr, V1460~Her and RX~J06480-4418) and one AXP~(4U~0142+61) described as a WD pulsar. We applied the matter accretion mechanism along with the magnetic deformation mechanism to evaluate the influence that the asymmetry caused by one process has on the deformability caused by the other process. We further calculated the GW amplitude for a mass range established by the inferred minimum and maximum mass self-consistently with rotation and use a realistic equation of state to explore the star's stability. The angle of inclination of the magnetic field in relation to the axis of rotation and the amount of mass accumulated in the magnetic poles are also parameters considered in the calculation of the amplitude.

We have shown that the effect of magnetic field deformation is negligible when we take into account the GW amplitude generated by the accretion mechanism for all sources. However, it is important to mention that if the WDs have stronger magnetic fields (AXP~4U~0142+61 described as a pulsar WD, for example, can have $B$ of up to $\sim 10^{10}$~G; see e.g., \cite{2012PASJ...64...56M, 2014PASJ...66...14C}), the effect produced by the field can become relevant by creating regions where the magnetic deformation mechanism is dominant over the accretion process. These regions are defined by the mass accumulation value $\delta m$ and the WD mass.

We have also inferred the mass and radius limits from the stability conditions of the WDs, where the minimum mass is obtained assuming that the observed rotation period corresponds to the Kepler period, that is, the WD rotates close to the mass-shedding limit. The maximum mass, in turn, is calculated from the occurrence of pycnonuclear reactions. Thus, we found that the analyzed WDs are possible GW sources for BBO and DECIGO given a certain combination of parameters, such as mass, inclination between the magnetic and rotational axes, and the accumulation of mass at the magnetic poles. However, this class of stars is unlikely to be detected by the LISA and TianQin space detectors due to a difficult combination of parameters such as a large accumulated mass at their magnetic poles, a large angle between the magnetic and rotational axes and a small WD mass value close to $M_{\rm min}$.

In addition, it is worth stressing the advantages of the detailed WD modeling that takes rotation and a realistic EoS into account, as it made it possible to obtain a range of mass/radius values that are possible for the analyzed sources. With the application of this model, it was possible to estimate constraints on the GW amplitude to verify the detectability of these WDs for different values of mass/radius, since these parameters are not yet well known.

%%%%%%%%%%%%%%%%%%%%%%%%%%%%%%%%%%%%%%%%%%%%%%%%%%%%%%%%%%%%%%%%%%%%%%%
%%%%%%%%%%%%%%%%%%%%%%%%%%%%%%%%%%%%%%%%%%%%%%%%%%%%%%%%%%%%%%%%%%%%%%%
\section*{Acknowledgements}
%%%%%%%%%%%%%%%%%%%%%%%%%%%%%%%%%%%%%%%%%%%%%%%%%%%%%%%%%%%%%%%%%%%%%%%
%%%%%%%%%%%%%%%%%%%%%%%%%%%%%%%%%%%%%%%%%%%%%%%%%%%%%%%%%%%%%%%%%%%%%%%

We thank the anonymous referees for the relevant suggestions which helped improve this work. M.F.S. and J.C.N.A. thank the Brazilian Ministry of Science, Technology and Innovation (MCTI) and the Brazilian Space Agency (AEB) who supported the present work under the PO 20VB.0009. M.F.S. thanks CAPES and Fundação Araucária for the financial support. J.G.C. is grateful for the support of FAPES (1020/2022, 1081/2022, 976/2022, 332/2023), CNPq (311758/2021-5), and FAPESP (2021/01089-1). M.F.S. and J.G.C. gratefully acknowledge the financial support of the ``Fen\^omenos Extremos do Universo" of the Funda\c{c}\~ao Arauc\'aria. E. Otoniel thanks for the support received from the productivity scholarship program, encouragement of internalization and technological innovation - BPI, notice (BP5-0197-00153.01.00/22). J.C.N.A. thanks CNPq (307803/2022-8) for partial financial support.

%%%%%%%%%%%%%%%%%%%%%%%%%%%%%%%%%%%%%%%%%%%%%%%%%%%%%%%%%%%%%%%%%%%%%%%
%%%%%%%%%%%%%%%%%%%%%%%%%%%%%%%%%%%%%%%%%%%%%%%%%%%%%%%%%%%%%%%%%%%%%%%
\section*{Data Availability}
%%%%%%%%%%%%%%%%%%%%%%%%%%%%%%%%%%%%%%%%%%%%%%%%%%%%%%%%%%%%%%%%%%%%%%%
%%%%%%%%%%%%%%%%%%%%%%%%%%%%%%%%%%%%%%%%%%%%%%%%%%%%%%%%%%%%%%%%%%%%%%%
 
The data underlying this article will be shared on reasonable request to the corresponding author.



\bibliographystyle{mnras}
\bibliography{reference} % if your bibtex file is called example.bib

\begin{thebibliography}{}
\makeatletter
\relax
\def\mn@urlcharsother{\let\do\@makeother \do\$\do\&\do\#\do\^\do\_\do\%\do\~}
\def\mn@doi{\begingroup\mn@urlcharsother \@ifnextchar [ {\mn@doi@} {\mn@doi@[]}}
\def\mn@doi@[#1]#2{\def\@tempa{#1}\ifx\@tempa\@empty \href {http://dx.doi.org/#2} {doi:#2}\else \href {http://dx.doi.org/#2} {#1}\fi \endgroup}
\def\mn@eprint#1#2{\mn@eprint@#1:#2::\@nil}
\def\mn@eprint@arXiv#1{\href {http://arxiv.org/abs/#1} {{\tt arXiv:#1}}}
\def\mn@eprint@dblp#1{\href {http://dblp.uni-trier.de/rec/bibtex/#1.xml} {dblp:#1}}
\def\mn@eprint@#1:#2:#3:#4\@nil{\def\@tempa {#1}\def\@tempb {#2}\def\@tempc {#3}\ifx \@tempc \@empty \let \@tempc \@tempb \let \@tempb \@tempa \fi \ifx \@tempb \@empty \def\@tempb {arXiv}\fi \@ifundefined {mn@eprint@\@tempb}{\@tempb:\@tempc}{\expandafter \expandafter \csname mn@eprint@\@tempb\endcsname \expandafter{\@tempc}}}

\bibitem[\protect\citeauthoryear{Abbott et~al.}{Abbott et~al.}{2016}]{2016PhRvL.116f1102A}
Abbott B.~P.,  et~al., 2016, \mn@doi [\prl] {10.1103/PhysRevLett.116.061102}, \href {http://adsabs.harvard.edu/abs/2016PhRvL.116f1102A} {116, 061102}

\bibitem[\protect\citeauthoryear{{Abbott} et~al.}{{Abbott} et~al.}{2019}]{2019PhRvX...9c1040A}
{Abbott} B.~P.,  et~al., 2019, \mn@doi [Physical Review X] {10.1103/PhysRevX.9.031040}, \href {https://ui.adsabs.harvard.edu/abs/2019PhRvX...9c1040A} {9, 031040}

\bibitem[\protect\citeauthoryear{{Abbott} et~al.}{{Abbott} et~al.}{2021}]{2021PhRvX..11b1053A}
{Abbott} R.,  et~al., 2021, \mn@doi [Physical Review X] {10.1103/PhysRevX.11.021053}, \href {https://ui.adsabs.harvard.edu/abs/2021PhRvX..11b1053A} {11, 021053}

\bibitem[\protect\citeauthoryear{{Althaus}, {Garc{\'\i}a-Berro}, {Isern}  \& {C{\'o}rsico}}{{Althaus} et~al.}{2005}]{2005A&A...441..689A}
{Althaus} L.~G.,  {Garc{\'\i}a-Berro} E.,  {Isern} J.,   {C{\'o}rsico} A.~H.,  2005, \mn@doi [\aap] {10.1051/0004-6361:20052996}, \href {https://ui.adsabs.harvard.edu/abs/2005A&A...441..689A} {441, 689}

\bibitem[\protect\citeauthoryear{{Althaus}, {Garc{\'\i}a-Berro}, {Isern}, {C{\'o}rsico}  \& {Rohrmann}}{{Althaus} et~al.}{2007}]{2007A&A...465..249A}
{Althaus} L.~G.,  {Garc{\'\i}a-Berro} E.,  {Isern} J.,  {C{\'o}rsico} A.~H.,   {Rohrmann} R.~D.,  2007, \mn@doi [\aap] {10.1051/0004-6361:20066059}, \href {https://ui.adsabs.harvard.edu/abs/2007A&A...465..249A} {465, 249}

\bibitem[\protect\citeauthoryear{Amaro-Seoane et~al.}{Amaro-Seoane et~al.}{2017}]{AMARO/2017}
Amaro-Seoane P.,  et~al., 2017, preprint (\mn@eprint {arXiv} {1702.00786})

\bibitem[\protect\citeauthoryear{{Amorim}, {Kepler}, {K{\"u}lebi}, {Jordan}  \& {Romero}}{{Amorim} et~al.}{2023}]{2023ApJ...944...56A}
{Amorim} L.~L.,  {Kepler} S.~O.,  {K{\"u}lebi} B.,  {Jordan} S.,   {Romero} A.~D.,  2023, \mn@doi [\apj] {10.3847/1538-4357/acaf6e}, \href {https://ui.adsabs.harvard.edu/abs/2023ApJ...944...56A} {944, 56}

\bibitem[\protect\citeauthoryear{{Ashley} et~al.,}{{Ashley} et~al.}{2020}]{Ashley2020}
{Ashley} R.~P.,  et~al., 2020, \mn@doi [\mnras] {10.1093/mnras/staa2676}, \href {https://ui.adsabs.harvard.edu/abs/2020MNRAS.499..149A} {499, 149}

\bibitem[\protect\citeauthoryear{{Barstow}, {Jordan}, {O'Donoghue}, {Burleigh}, {Napiwotzki}  \& {Harrop-Allin}}{{Barstow} et~al.}{1995}]{1995MNRAS.277..971B}
{Barstow} M.~A.,  {Jordan} S.,  {O'Donoghue} D.,  {Burleigh} M.~R.,  {Napiwotzki} R.,   {Harrop-Allin} M.~K.,  1995, \mnras, \href {http://adsabs.harvard.edu/abs/1995MNRAS.277..971B} {277, 971}

\bibitem[\protect\citeauthoryear{{Becerra}, {Rueda}, {Lor{\'e}n-Aguilar}  \& {Garc{\'\i}a-Berro}}{{Becerra} et~al.}{2018}]{2018ApJ...857..134B}
{Becerra} L.,  {Rueda} J.~A.,  {Lor{\'e}n-Aguilar} P.,   {Garc{\'\i}a-Berro} E.,  2018, \mn@doi [\apj] {10.3847/1538-4357/aabc12}, \href {https://ui.adsabs.harvard.edu/abs/2018ApJ...857..134B} {857, 134}

\bibitem[\protect\citeauthoryear{{Bonazzola} \& {Gourgoulhon}}{{Bonazzola} \& {Gourgoulhon}}{1996}]{1996A&A...312..675B}
{Bonazzola} S.,  {Gourgoulhon} E.,  1996, \aap, \href {https://ui.adsabs.harvard.edu/abs/1996A\%26A...312..675B} {312, 675}

\bibitem[\protect\citeauthoryear{{Borges}, {Rodrigues}, {Coelho}, {Malheiro}  \& {Castro}}{{Borges} et~al.}{2020}]{Borges2020}
{Borges} S.~V.,  {Rodrigues} C.~V.,  {Coelho} J.~G.,  {Malheiro} M.,   {Castro} M.,  2020, \mn@doi [\apj] {10.3847/1538-4357/ab8add}, \href {https://ui.adsabs.harvard.edu/abs/2020ApJ...895...26B} {895, 26}

\bibitem[\protect\citeauthoryear{Boshkayev, Rueda  \& Ruffini}{Boshkayev et~al.}{2013a}]{boshkayev_maximum_2013}
Boshkayev K.,  Rueda J.,   Ruffini R.,  2013a, \mn@doi [International Journal of Modern Physics: Conference Series] {10.1142/S201019451301129X}, 23, 193

\bibitem[\protect\citeauthoryear{{Boshkayev}, {Rueda}, {Ruffini}  \& {Siutsou}}{{Boshkayev} et~al.}{2013b}]{2013ApJ...762..117B}
{Boshkayev} K.,  {Rueda} J.~A.,  {Ruffini} R.,   {Siutsou} I.,  2013b, \mn@doi [\apj] {10.1088/0004-637X/762/2/117}, \href {https://ui.adsabs.harvard.edu/abs/2013ApJ...762..117B} {762, 117}

\bibitem[\protect\citeauthoryear{{Buckley}, {Meintjes}, {Potter}, {Marsh}  \& {G{\"a}nsicke}}{{Buckley} et~al.}{2017}]{2017NatAs...1E..29B}
{Buckley} D.~A.~H.,  {Meintjes} P.~J.,  {Potter} S.~B.,  {Marsh} T.~R.,   {G{\"a}nsicke} B.~T.,  2017, \mn@doi [Nature Astronomy] {10.1038/s41550-016-0029}, \href {https://ui.adsabs.harvard.edu/abs/2017NatAs...1E..29B} {1, 0029}

\bibitem[\protect\citeauthoryear{{Caiazzo} et~al.,}{{Caiazzo} et~al.}{2021}]{2021Natur.595...39C}
{Caiazzo} I.,  et~al., 2021, \mn@doi [\nat] {10.1038/s41586-021-03615-y}, \href {https://ui.adsabs.harvard.edu/abs/2021Natur.595...39C} {595, 39}

\bibitem[\protect\citeauthoryear{{Camisassa} et~al.,}{{Camisassa} et~al.}{2019}]{2019A&A...625A..87C}
{Camisassa} M.~E.,  et~al., 2019, \mn@doi [\aap] {10.1051/0004-6361/201833822}, \href {https://ui.adsabs.harvard.edu/abs/2019A&A...625A..87C} {625, A87}

\bibitem[\protect\citeauthoryear{Chabrier \& Potekhin}{Chabrier \& Potekhin}{1998}]{PhysRevE.58.4941}
Chabrier G.,  Potekhin A.~Y.,  1998, \mn@doi [Phys. Rev. E] {10.1103/PhysRevE.58.4941}, 58, 4941

\bibitem[\protect\citeauthoryear{{Chamel}, {Fantina}  \& {Davis}}{{Chamel} et~al.}{2013}]{2013PhRvD..88h1301C}
{Chamel} N.,  {Fantina} A.~F.,   {Davis} P.~J.,  2013, \mn@doi [\prd] {10.1103/PhysRevD.88.081301}, \href {https://ui.adsabs.harvard.edu/abs/2013PhRvD..88h1301C} {88, 081301}

\bibitem[\protect\citeauthoryear{Chamel, Molter, Fantina  \& Arteaga}{Chamel et~al.}{2014}]{chamel_maximum_2014}
Chamel N.,  Molter E.,  Fantina A.,   Arteaga D.~P.,  2014, \prd, 90, 043002

\bibitem[\protect\citeauthoryear{{Chatterjee}, {Fantina}, {Chamel}, {Novak}  \& {Oertel}}{{Chatterjee} et~al.}{2017}]{2017MNRAS.469...95C}
{Chatterjee} D.,  {Fantina} A.~F.,  {Chamel} N.,  {Novak} J.,   {Oertel} M.,  2017, \mn@doi [\mnras] {10.1093/mnras/stx781}, \href {https://ui.adsabs.harvard.edu/abs/2017MNRAS.469...95C} {469, 95}

\bibitem[\protect\citeauthoryear{{Coelho} \& {Malheiro}}{{Coelho} \& {Malheiro}}{2014}]{2014PASJ...66...14C}
{Coelho} J.~G.,  {Malheiro} M.,  2014, \mn@doi [\pasj] {10.1093/pasj/pst014}, \href {https://ui.adsabs.harvard.edu/abs/2014PASJ...66...14C} {66, 14}

\bibitem[\protect\citeauthoryear{{Coelho}, {Marinho}, {Malheiro}, {Negreiros}, {C{\'a}ceres}, {Rueda}  \& {Ruffini}}{{Coelho} et~al.}{2014}]{2014ApJ...794...86C}
{Coelho} J.~G.,  {Marinho} R.~M.,  {Malheiro} M.,  {Negreiros} R.,  {C{\'a}ceres} D.~L.,  {Rueda} J.~A.,   {Ruffini} R.,  2014, \mn@doi [\apj] {10.1088/0004-637X/794/1/86}, \href {https://ui.adsabs.harvard.edu/abs/2014ApJ...794...86C} {794, 86}

\bibitem[\protect\citeauthoryear{{Coelho}, {C{\'a}ceres}, {de Lima}, {Malheiro}, {Rueda}  \& {Ruffini}}{{Coelho} et~al.}{2017}]{2017A&A...599A..87C}
{Coelho} J.~G.,  {C{\'a}ceres} D.~L.,  {de Lima} R.~C.~R.,  {Malheiro} M.,  {Rueda} J.~A.,   {Ruffini} R.,  2017, \mn@doi [\aap] {10.1051/0004-6361/201629521}, \href {https://ui.adsabs.harvard.edu/abs/2017A&A...599A..87C} {599, A87}

\bibitem[\protect\citeauthoryear{{C{\'o}rsico}, {Althaus}, {Miller Bertolami}  \& {Kepler}}{{C{\'o}rsico} et~al.}{2019}]{2019A&ARv..27....7C}
{C{\'o}rsico} A.~H.,  {Althaus} L.~G.,  {Miller Bertolami} M.~M.,   {Kepler} S.~O.,  2019, \mn@doi [\aapr] {10.1007/s00159-019-0118-4}, \href {https://ui.adsabs.harvard.edu/abs/2019A&ARv..27....7C} {27, 7}

\bibitem[\protect\citeauthoryear{{Curd}, {Gianninas}, {Bell}, {Kilic}, {Romero}, {Allende Prieto}, {Winget}  \& {Winget}}{{Curd} et~al.}{2017}]{2017MNRAS.468..239C}
{Curd} B.,  {Gianninas} A.,  {Bell} K.~J.,  {Kilic} M.,  {Romero} A.~D.,  {Allende Prieto} C.,  {Winget} D.~E.,   {Winget} K.~I.,  2017, \mn@doi [\mnras] {10.1093/mnras/stx320}, \href {https://ui.adsabs.harvard.edu/abs/2017MNRAS.468..239C} {468, 239}

\bibitem[\protect\citeauthoryear{{Dan}, {Rosswog}, {Br{\"u}ggen}  \& {Podsiadlowski}}{{Dan} et~al.}{2014}]{2014MNRAS.438...14D}
{Dan} M.,  {Rosswog} S.,  {Br{\"u}ggen} M.,   {Podsiadlowski} P.,  2014, \mn@doi [\mnras] {10.1093/mnras/stt1766}, \href {https://ui.adsabs.harvard.edu/abs/2014MNRAS.438...14D} {438, 14}

\bibitem[\protect\citeauthoryear{Eisenstein et~al.,}{Eisenstein et~al.}{2006}]{eisenstein_catalog_2006}
Eisenstein D.~J.,  et~al., 2006, The Astrophysical Journal Supplement Series, 167, 40

\bibitem[\protect\citeauthoryear{{Ferrario}, {de Martino}  \& {G{\"a}nsicke}}{{Ferrario} et~al.}{2015}]{2015SSRv..191..111F}
{Ferrario} L.,  {de Martino} D.,   {G{\"a}nsicke} B.~T.,  2015, \mn@doi [\ssr] {10.1007/s11214-015-0152-0}, \href {https://ui.adsabs.harvard.edu/abs/2015SSRv..191..111F} {191, 111}

\bibitem[\protect\citeauthoryear{{Ferrario}, {Wickramasinghe}  \& {Kawka}}{{Ferrario} et~al.}{2020}]{2020AdSpR..66.1025F}
{Ferrario} L.,  {Wickramasinghe} D.,   {Kawka} A.,  2020, \mn@doi [Advances in Space Research] {10.1016/j.asr.2019.11.012}, \href {https://ui.adsabs.harvard.edu/abs/2020AdSpR..66.1025F} {66, 1025}

\bibitem[\protect\citeauthoryear{{Ferraro}}{{Ferraro}}{1954}]{1954ApJ...119..407F}
{Ferraro} V.~C.~A.,  1954, \mn@doi [\apj] {10.1086/145838}, \href {https://ui.adsabs.harvard.edu/abs/1954ApJ...119..407F} {119, 407}

\bibitem[\protect\citeauthoryear{{Franzon} \& {Schramm}}{{Franzon} \& {Schramm}}{2017}]{2017MNRAS.467.4484F}
{Franzon} B.,  {Schramm} S.,  2017, \mn@doi [\mnras] {10.1093/mnras/stx397}, \href {https://ui.adsabs.harvard.edu/abs/2017MNRAS.467.4484F} {467, 4484}

\bibitem[\protect\citeauthoryear{{Friedman}, {Ipser}  \& {Parker}}{{Friedman} et~al.}{1986}]{Friedman1986}
{Friedman} J.~L.,  {Ipser} J.~R.,   {Parker} L.,  1986, \mn@doi [\apj] {10.1086/164149}, \href {https://ui.adsabs.harvard.edu/abs/1986ApJ...304..115F} {304, 115}

\bibitem[\protect\citeauthoryear{Gaia{ }Collaboration, Vallenari, Brown, Prusti, de{ }Bruijne  \& Arenou}{Gaia{ }Collaboration et~al.}{2022}]{Gaia2022}
Gaia{ }Collaboration Vallenari A.,  Brown A.,  Prusti T.,  de{ }Bruijne J.,   Arenou F.,  2022, Astronomy \& Astrophysics

\bibitem[\protect\citeauthoryear{{Gao}, {Cao}  \& {Zhang}}{{Gao} et~al.}{2017}]{2017ApJ...844..112G}
{Gao} H.,  {Cao} Z.,   {Zhang} B.,  2017, \mn@doi [\apj] {10.3847/1538-4357/aa7d00}, \href {https://ui.adsabs.harvard.edu/abs/2017ApJ...844..112G} {844, 112}

\bibitem[\protect\citeauthoryear{{Garc{\'\i}a-Berro} et~al.,}{{Garc{\'\i}a-Berro} et~al.}{2012}]{2012ApJ...749...25G}
{Garc{\'\i}a-Berro} E.,  et~al., 2012, \mn@doi [\apj] {10.1088/0004-637X/749/1/25}, \href {https://ui.adsabs.harvard.edu/abs/2012ApJ...749...25G} {749, 25}

\bibitem[\protect\citeauthoryear{{Gasques}, {Afanasjev}, {Aguilera}, {Beard}, {Chamon}, {Ring}, {Wiescher}  \& {Yakovlev}}{{Gasques} et~al.}{2005}]{2005PhRvC..72b5806G}
{Gasques} L.~R.,  {Afanasjev} A.~V.,  {Aguilera} E.~F.,  {Beard} M.,  {Chamon} L.~C.,  {Ring} P.,  {Wiescher} M.,   {Yakovlev} D.~G.,  2005, \mn@doi [\prc] {10.1103/PhysRevC.72.025806}, \href {https://ui.adsabs.harvard.edu/abs/2005PhRvC..72b5806G} {72, 025806}

\bibitem[\protect\citeauthoryear{Gentile~Fusillo et~al.,}{Gentile~Fusillo et~al.}{2018}]{10.1093/mnras/sty3016}
Gentile~Fusillo N.~P.,  et~al., 2018, \mn@doi [Monthly Notices of the Royal Astronomical Society] {10.1093/mnras/sty3016}, 482, 4570

\bibitem[\protect\citeauthoryear{Glendenning \& Weber}{Glendenning \& Weber}{1994}]{PhysRevD.50.3836}
Glendenning N.~K.,  Weber F.,  1994, \mn@doi [Phys. Rev. D] {10.1103/PhysRevD.50.3836}, 50, 3836

\bibitem[\protect\citeauthoryear{{Golf}, {Hellmers}  \& {Weber}}{{Golf} et~al.}{2009}]{2009PhRvC..80a5804G}
{Golf} B.,  {Hellmers} J.,   {Weber} F.,  2009, \mn@doi [\prc] {10.1103/PhysRevC.80.015804}, \href {https://ui.adsabs.harvard.edu/abs/2009PhRvC..80a5804G} {80, 015804}

\bibitem[\protect\citeauthoryear{{Harry}, {Fritschel}, {Shaddock}, {Folkner}  \& {Phinney}}{{Harry} et~al.}{2006}]{2006CQGra..23.4887H}
{Harry} G.~M.,  {Fritschel} P.,  {Shaddock} D.~A.,  {Folkner} W.,   {Phinney} E.~S.,  2006, \mn@doi [Classical and Quantum Gravity] {10.1088/0264-9381/23/15/008}, \href {https://ui.adsabs.harvard.edu/abs/2006CQGra..23.4887H} {23, 4887}

\bibitem[\protect\citeauthoryear{{Hartle}}{{Hartle}}{1967}]{1967ApJ...150.1005H}
{Hartle} J.~B.,  1967, \mn@doi [\apj] {10.1086/149400}, \href {http://adsabs.harvard.edu/abs/1967ApJ...150.1005H} {150, 1005}

\bibitem[\protect\citeauthoryear{{Hartle} \& {Thorne}}{{Hartle} \& {Thorne}}{1968}]{1968ApJ...153..807H}
{Hartle} J.~B.,  {Thorne} K.~S.,  1968, \mn@doi [\apj] {10.1086/149707}, \href {http://adsabs.harvard.edu/abs/1968ApJ...153..807H} {153, 807}

\bibitem[\protect\citeauthoryear{{Hermes}, {Kepler}, {Castanheira}, {Gianninas}, {Winget}, {Montgomery}, {Brown}  \& {Harrold}}{{Hermes} et~al.}{2013}]{2013ApJ...771L...2H}
{Hermes} J.~J.,  {Kepler} S.~O.,  {Castanheira} B.~G.,  {Gianninas} A.,  {Winget} D.~E.,  {Montgomery} M.~H.,  {Brown} W.~R.,   {Harrold} S.~T.,  2013, \mn@doi [\apjl] {10.1088/2041-8205/771/1/L2}, \href {https://ui.adsabs.harvard.edu/abs/2013ApJ...771L...2H} {771, L2}

\bibitem[\protect\citeauthoryear{{Israel}, {Stella}, {Angelini}, {White}, {Kallman}, {Giommi}  \& {Treves}}{{Israel} et~al.}{1997}]{Israel1997}
{Israel} G.~L.,  {Stella} L.,  {Angelini} L.,  {White} N.~E.,  {Kallman} T.~R.,  {Giommi} P.,   {Treves} A.,  1997, \mn@doi [\apjl] {10.1086/310418}, \href {https://ui.adsabs.harvard.edu/abs/1997ApJ...474L..53I} {474, L53}

\bibitem[\protect\citeauthoryear{{Jim{\'e}nez-Esteban}, {Torres}, {Rebassa-Mansergas}, {Skorobogatov}, {Solano}, {Cantero}  \& {Rodrigo}}{{Jim{\'e}nez-Esteban} et~al.}{2018}]{2018MNRAS.480.4505J}
{Jim{\'e}nez-Esteban} F.~M.,  {Torres} S.,  {Rebassa-Mansergas} A.,  {Skorobogatov} G.,  {Solano} E.,  {Cantero} C.,   {Rodrigo} C.,  2018, \mn@doi [\mnras] {10.1093/mnras/sty2120}, \href {https://ui.adsabs.harvard.edu/abs/2018MNRAS.480.4505J} {480, 4505}

\bibitem[\protect\citeauthoryear{{Jos{\'e}}, {Shore}  \& {Casanova}}{{Jos{\'e}} et~al.}{2020}]{2020A&A...634A...5J}
{Jos{\'e}} J.,  {Shore} S.~N.,   {Casanova} J.,  2020, \mn@doi [\aap] {10.1051/0004-6361/201936893}, \href {https://ui.adsabs.harvard.edu/abs/2020A&A...634A...5J} {634, A5}

\bibitem[\protect\citeauthoryear{{Kalita} \& {Mukhopadhyay}}{{Kalita} \& {Mukhopadhyay}}{2019}]{2019MNRAS.490.2692K}
{Kalita} S.,  {Mukhopadhyay} B.,  2019, \mn@doi [\mnras] {10.1093/mnras/stz2734}, \href {https://ui.adsabs.harvard.edu/abs/2019MNRAS.490.2692K} {490, 2692}

\bibitem[\protect\citeauthoryear{{Kalita}, {Mukhopadhyay}, {Mondal}  \& {Bulik}}{{Kalita} et~al.}{2020}]{2020ApJ...896...69K}
{Kalita} S.,  {Mukhopadhyay} B.,  {Mondal} T.,   {Bulik} T.,  2020, \mn@doi [\apj] {10.3847/1538-4357/ab8e40}, \href {https://ui.adsabs.harvard.edu/abs/2020ApJ...896...69K} {896, 69}

\bibitem[\protect\citeauthoryear{{Kawamura} et~al.,}{{Kawamura} et~al.}{2006}]{2006CQGra..23S.125K}
{Kawamura} S.,  et~al., 2006, \mn@doi [Classical and Quantum Gravity] {10.1088/0264-9381/23/8/S17}, \href {https://ui.adsabs.harvard.edu/abs/2006CQGra..23S.125K} {23, S125}

\bibitem[\protect\citeauthoryear{Kepler et~al.}{Kepler et~al.}{2013}]{kepler/2013}
Kepler S.~O.,  et~al., 2013, Monthly Notices of the Royal Astronomical Society, 429, 2934

\bibitem[\protect\citeauthoryear{{Kepler} et~al.,}{{Kepler} et~al.}{2015}]{2015MNRAS.446.4078K}
{Kepler} S.~O.,  et~al., 2015, \mn@doi [\mnras] {10.1093/mnras/stu2388}, \href {http://adsabs.harvard.edu/abs/2015MNRAS.446.4078K} {446, 4078}

\bibitem[\protect\citeauthoryear{{Kilic}, {Kosakowski}, {Moss}, {Bergeron}  \& {Conly}}{{Kilic} et~al.}{2021}]{2021ApJ...923L...6K}
{Kilic} M.,  {Kosakowski} A.,  {Moss} A.~G.,  {Bergeron} P.,   {Conly} A.~A.,  2021, \mn@doi [\apjl] {10.3847/2041-8213/ac3b60}, \href {https://ui.adsabs.harvard.edu/abs/2021ApJ...923L...6K} {923, L6}

\bibitem[\protect\citeauthoryear{{Kitaguchi} et~al.,}{{Kitaguchi} et~al.}{2014}]{Kitaguchi2014}
{Kitaguchi} T.,  et~al., 2014, \mn@doi [\apj] {10.1088/0004-637X/782/1/3}, \href {https://ui.adsabs.harvard.edu/abs/2014ApJ...782....3K} {782, 3}

\bibitem[\protect\citeauthoryear{Kjurkchieva, Popov, Vasileva  \& Petrov}{Kjurkchieva et~al.}{2017}]{KJURKCHIEVA2017}
Kjurkchieva D.~P.,  Popov V.~A.,  Vasileva D.~L.,   Petrov N.~I.,  2017, \mn@doi [New Astronomy] {https://doi.org/10.1016/j.newast.2016.10.001}, 52, 8

\bibitem[\protect\citeauthoryear{{Konno}, {Obata}  \& {Kojima}}{{Konno} et~al.}{2000}]{2000A&A...356..234K}
{Konno} K.,  {Obata} T.,   {Kojima} Y.,  2000, \aap, \href {http://adsabs.harvard.edu/abs/2000A\%26A...356..234K} {356, 234}

\bibitem[\protect\citeauthoryear{{Kuelebi}, {Jordan}, {Euchner}, {Gaensicke}  \& {Hirsch}}{{Kuelebi} et~al.}{2010}]{2010yCat..35061341K}
{Kuelebi} B.,  {Jordan} S.,  {Euchner} F.,  {Gaensicke} B.~T.,   {Hirsch} H.,  2010, VizieR Online Data Catalog, \href {https://ui.adsabs.harvard.edu/abs/2010yCat..35061341K} {pp J/A+A/506/1341}

\bibitem[\protect\citeauthoryear{{K{\"u}lebi}, {Jordan}, {Euchner}, {G{\"a}nsicke}  \& {Hirsch}}{{K{\"u}lebi} et~al.}{2009}]{2009A&A...506.1341K}
{K{\"u}lebi} B.,  {Jordan} S.,  {Euchner} F.,  {G{\"a}nsicke} B.~T.,   {Hirsch} H.,  2009, \mn@doi [\aap] {10.1051/0004-6361/200912570}, \href {https://ui.adsabs.harvard.edu/abs/2009A&A...506.1341K} {506, 1341}

\bibitem[\protect\citeauthoryear{{K{\"u}lebi}, {Jordan}, {Nelan}, {Bastian}  \& {Altmann}}{{K{\"u}lebi} et~al.}{2010}]{2010A&A...524A..36K}
{K{\"u}lebi} B.,  {Jordan} S.,  {Nelan} E.,  {Bastian} U.,   {Altmann} M.,  2010, \mn@doi [\aap] {10.1051/0004-6361/201015237}, \href {http://adsabs.harvard.edu/abs/2010A%26A...524A..36K} {524, A36}

\bibitem[\protect\citeauthoryear{{Lopes de Oliveira}, {Bruch}, {Rodrigues}, {Oliveira}  \& {Mukai}}{{Lopes de Oliveira} et~al.}{2020}]{LopesDeOliveira2020}
{Lopes de Oliveira} R.,  {Bruch} A.,  {Rodrigues} C.~V.,  {Oliveira} A.~S.,   {Mukai} K.,  2020, \mn@doi [\apjl] {10.3847/2041-8213/aba618}, \href {https://ui.adsabs.harvard.edu/abs/2020ApJ...898L..40L} {898, L40}

\bibitem[\protect\citeauthoryear{{Lor{\'e}n-Aguilar}, {Isern}  \& {Garc{\'\i}a-Berro}}{{Lor{\'e}n-Aguilar} et~al.}{2009}]{2009A&A...500.1193L}
{Lor{\'e}n-Aguilar} P.,  {Isern} J.,   {Garc{\'\i}a-Berro} E.,  2009, \mn@doi [\aap] {10.1051/0004-6361/200811060}, \href {https://ui.adsabs.harvard.edu/abs/2009A&A...500.1193L} {500, 1193}

\bibitem[\protect\citeauthoryear{{Lu}, {Tan}  \& {Shao}}{{Lu} et~al.}{2019}]{2019PhRvD.100d4042L}
{Lu} X.-Y.,  {Tan} Y.-J.,   {Shao} C.-G.,  2019, \mn@doi [\prd] {10.1103/PhysRevD.100.044042}, \href {https://ui.adsabs.harvard.edu/abs/2019PhRvD.100d4042L} {100, 044042}

\bibitem[\protect\citeauthoryear{{Luo} et~al.,}{{Luo} et~al.}{2016}]{2016CQGra..33c5010L}
{Luo} J.,  et~al., 2016, \mn@doi [Classical and Quantum Gravity] {10.1088/0264-9381/33/3/035010}, \href {https://ui.adsabs.harvard.edu/abs/2016CQGra..33c5010L} {33, 035010}

\bibitem[\protect\citeauthoryear{Maggiore}{Maggiore}{2008}]{maggiore/2008}
Maggiore M.,  2008, Gravitational waves: volume 1: theory and experiments.
Gravitational Waves, OUP Oxford

\bibitem[\protect\citeauthoryear{{Malheiro}, {Rueda}  \& {Ruffini}}{{Malheiro} et~al.}{2012}]{2012PASJ...64...56M}
{Malheiro} M.,  {Rueda} J.~A.,   {Ruffini} R.,  2012, \mn@doi [\pasj] {10.1093/pasj/64.3.56}, \href {https://ui.adsabs.harvard.edu/abs/2012PASJ...64...56M} {64, 56}

\bibitem[\protect\citeauthoryear{{Marsh} et~al.,}{{Marsh} et~al.}{2016}]{2016Natur.537..374M}
{Marsh} T.~R.,  et~al., 2016, \mn@doi [\nat] {10.1038/nature18620}, \href {https://ui.adsabs.harvard.edu/abs/2016Natur.537..374M} {537, 374}

\bibitem[\protect\citeauthoryear{{Mei} et~al.,}{{Mei} et~al.}{2021}]{2021PTEP.2021eA107M}
{Mei} J.,  et~al., 2021, \mn@doi [Progress of Theoretical and Experimental Physics] {10.1093/ptep/ptaa114}, \href {https://ui.adsabs.harvard.edu/abs/2021PTEP.2021eA107M} {2021, 05A107}

\bibitem[\protect\citeauthoryear{{Mereghetti}, {Tiengo}, {Esposito}, {La Palombara}, {Israel}  \& {Stella}}{{Mereghetti} et~al.}{2009}]{Mereghetti2009}
{Mereghetti} S.,  {Tiengo} A.,  {Esposito} P.,  {La Palombara} N.,  {Israel} G.~L.,   {Stella} L.,  2009, \mn@doi [Science] {10.1126/science.1176252}, \href {https://ui.adsabs.harvard.edu/abs/2009Sci...325.1222M} {325, 1222}

\bibitem[\protect\citeauthoryear{{Mereghetti}, {La Palombara}, {Tiengo}, {Pizzolato}, {Esposito}, {Woudt}, {Israel}  \& {Stella}}{{Mereghetti} et~al.}{2011}]{Mereghetti2011}
{Mereghetti} S.,  {La Palombara} N.,  {Tiengo} A.,  {Pizzolato} F.,  {Esposito} P.,  {Woudt} P.~A.,  {Israel} G.~L.,   {Stella} L.,  2011, \mn@doi [\apj] {10.1088/0004-637X/737/2/51}, \href {https://ui.adsabs.harvard.edu/abs/2011ApJ...737...51M} {737, 51}

\bibitem[\protect\citeauthoryear{{Mereghetti}, {Pintore}, {Esposito}, {La Palombara}, {Tiengo}, {Israel}  \& {Stella}}{{Mereghetti} et~al.}{2016}]{2016MNRAS.458.3523M}
{Mereghetti} S.,  {Pintore} F.,  {Esposito} P.,  {La Palombara} N.,  {Tiengo} A.,  {Israel} G.~L.,   {Stella} L.,  2016, \mn@doi [\mnras] {10.1093/mnras/stw536}, \href {https://ui.adsabs.harvard.edu/abs/2016MNRAS.458.3523M} {458, 3523}

\bibitem[\protect\citeauthoryear{{Mereghetti} et~al.,}{{Mereghetti} et~al.}{2021}]{Mereghetti2021}
{Mereghetti} S.,  et~al., 2021, \mn@doi [\mnras] {10.1093/mnras/stab1004}, \href {https://ui.adsabs.harvard.edu/abs/2021MNRAS.504..920M} {504, 920}

\bibitem[\protect\citeauthoryear{{Mitrofanov}}{{Mitrofanov}}{1980}]{1980IAUS...88..431M}
{Mitrofanov} I.~G.,  1980, in Close Binary Stars: Observations and Interpretation. pp 431--436

\bibitem[\protect\citeauthoryear{{Mukhopadhyay} \& {Bhattacharya}}{{Mukhopadhyay} \& {Bhattacharya}}{2022}]{2022Parti...5..493M}
{Mukhopadhyay} B.,  {Bhattacharya} M.,  2022, \mn@doi [Particles] {10.3390/particles5040037}, \href {https://ui.adsabs.harvard.edu/abs/2022Parti...5..493M} {5, 493}

\bibitem[\protect\citeauthoryear{{Mukhopadhyay} \& {Rao}}{{Mukhopadhyay} \& {Rao}}{2016}]{2016JCAP...05..007M}
{Mukhopadhyay} B.,  {Rao} A.~R.,  2016, \mn@doi [\jcap] {10.1088/1475-7516/2016/05/007}, \href {https://ui.adsabs.harvard.edu/abs/2016JCAP...05..007M} {2016, 007}

\bibitem[\protect\citeauthoryear{{Mukhopadhyay}, {Rao}  \& {Bhatia}}{{Mukhopadhyay} et~al.}{2017}]{2017MNRAS.472.3564M}
{Mukhopadhyay} B.,  {Rao} A.~R.,   {Bhatia} T.~S.,  2017, \mn@doi [\mnras] {10.1093/mnras/stx2119}, \href {https://ui.adsabs.harvard.edu/abs/2017MNRAS.472.3564M} {472, 3564}

\bibitem[\protect\citeauthoryear{{Nunes}, {Arba{\~n}il}  \& {Malheiro}}{{Nunes} et~al.}{2021}]{2021ApJ...921..138N}
{Nunes} S.~P.,  {Arba{\~n}il} J. D.~V.,   {Malheiro} M.,  2021, \mn@doi [\apj] {10.3847/1538-4357/ac1e8a}, \href {https://ui.adsabs.harvard.edu/abs/2021ApJ...921..138N} {921, 138}

\bibitem[\protect\citeauthoryear{{Olausen} \& {Kaspi}}{{Olausen} \& {Kaspi}}{2014}]{Olausen&Kaspi2014}
{Olausen} S.~A.,  {Kaspi} V.~M.,  2014, \mn@doi [\apjs] {10.1088/0067-0049/212/1/6}, \href {https://ui.adsabs.harvard.edu/abs/2014ApJS..212....6O} {212, 6}

\bibitem[\protect\citeauthoryear{Otoniel, Franzon, Carvalho, Malheiro, Schramm  \& Weber}{Otoniel et~al.}{2019}]{otoniel_strongly_2019}
Otoniel E.,  Franzon B.,  Carvalho G.~A.,  Malheiro M.,  Schramm S.,   Weber F.,  2019, \mn@doi [The Astrophysical Journal] {10.3847/1538-4357/ab24d1}, 879, 46

\bibitem[\protect\citeauthoryear{{Otoniel}, {Coelho}, {Nunes}, {Malheiro}  \& {Weber}}{{Otoniel} et~al.}{2021}]{Otoniel2021}
{Otoniel} E.,  {Coelho} J.~G.,  {Nunes} S.~P.,  {Malheiro} M.,   {Weber} F.,  2021, \mn@doi [\aap] {10.1051/0004-6361/202039749}, \href {https://ui.adsabs.harvard.edu/abs/2021A&A...656A..77O} {656, A77}

\bibitem[\protect\citeauthoryear{{Otoniel}, {Coelho}, {Malheiro}  \& {Weber}}{{Otoniel} et~al.}{2022}]{2022atcc.book..121O}
{Otoniel} E.,  {Coelho} J.~G.,  {Malheiro} M.,   {Weber} F.,  2022, in , Astrophysics in the XXI Century with Compact Stars. Edited by C.A.Z. Vasconcellos. eISBN 978-981-12-2094-4. Singapore: World Scientific.
World Scientific, pp 121--152, \mn@doi{10.1142/9789811220944_0004}

\bibitem[\protect\citeauthoryear{{Patterson}}{{Patterson}}{1979}]{1979ApJ...234..978P}
{Patterson} J.,  1979, \mn@doi [\apj] {10.1086/157582}, \href {https://ui.adsabs.harvard.edu/abs/1979ApJ...234..978P} {234, 978}

\bibitem[\protect\citeauthoryear{{Pelisoli} et~al.,}{{Pelisoli} et~al.}{2022}]{Pelisoli2022}
{Pelisoli} I.,  et~al., 2022, \mn@doi [\mnras] {10.1093/mnrasl/slab116}, \href {https://ui.adsabs.harvard.edu/abs/2022MNRAS.509L..31P} {509, L31}

\bibitem[\protect\citeauthoryear{{Pelisoli} et~al.,}{{Pelisoli} et~al.}{2023}]{2023NatAs.tmp..120P}
{Pelisoli} I.,  et~al., 2023, \mn@doi [Nature Astronomy] {10.1038/s41550-023-01995-x}, \href {https://ui.adsabs.harvard.edu/abs/2023NatAs.tmp..120P} {}

\bibitem[\protect\citeauthoryear{{Pereira}, {Coelho}  \& {de Lima}}{{Pereira} et~al.}{2018}]{2018EPJC...78..361P}
{Pereira} J.~P.,  {Coelho} J.~G.,   {de Lima} R. C.~R.,  2018, \mn@doi [European Physical Journal C] {10.1140/epjc/s10052-018-5849-2}, \href {https://ui.adsabs.harvard.edu/abs/2018EPJC...78..361P} {78, 361}

\bibitem[\protect\citeauthoryear{{Popov}, {Mereghetti}, {Blinnikov}, {Kuranov}  \& {Yungelson}}{{Popov} et~al.}{2018}]{2018MNRAS.474.2750P}
{Popov} S.~B.,  {Mereghetti} S.,  {Blinnikov} S.~I.,  {Kuranov} A.~G.,   {Yungelson} L.~R.,  2018, \mn@doi [\mnras] {10.1093/mnras/stx2910}, \href {https://ui.adsabs.harvard.edu/abs/2018MNRAS.474.2750P} {474, 2750}

\bibitem[\protect\citeauthoryear{Potekhin \& Chabrier}{Potekhin \& Chabrier}{2000}]{potekhin_equation_2000}
Potekhin A.~Y.,  Chabrier G.,  2000, \mn@doi [Physical Review E] {10.1103/PhysRevE.62.8554}, 62, 8554

\bibitem[\protect\citeauthoryear{{Regimbau} \& {de Freitas Pacheco}}{{Regimbau} \& {de Freitas Pacheco}}{2006}]{2006A&A...447....1R}
{Regimbau} T.,  {de Freitas Pacheco} J.~A.,  2006, \mn@doi [\aap] {10.1051/0004-6361:20053702}, \href {http://adsabs.harvard.edu/abs/2006A\%26A...447....1R} {447, 1}

\bibitem[\protect\citeauthoryear{{Robson}, {Cornish}  \& {Liu}}{{Robson} et~al.}{2019}]{2019CQGra..36j5011R}
{Robson} T.,  {Cornish} N.~J.,   {Liu} C.,  2019, \mn@doi [Classical and Quantum Gravity] {10.1088/1361-6382/ab1101}, \href {https://ui.adsabs.harvard.edu/abs/2019CQGra..36j5011R} {36, 105011}

\bibitem[\protect\citeauthoryear{{Rueda}, {Boshkayev}, {Izzo}, {Ruffini}, {Lor{\'e}n-Aguilar}, {K{\"u}lebi}, {Aznar-Sigu{\'a}n}  \& {Garc{\'\i}a-Berro}}{{Rueda} et~al.}{2013}]{2013ApJ...772L..24R}
{Rueda} J.~A.,  {Boshkayev} K.,  {Izzo} L.,  {Ruffini} R.,  {Lor{\'e}n-Aguilar} P.,  {K{\"u}lebi} B.,  {Aznar-Sigu{\'a}n} G.,   {Garc{\'\i}a-Berro} E.,  2013, \mn@doi [\apjl] {10.1088/2041-8205/772/2/L24}, \href {https://ui.adsabs.harvard.edu/abs/2013ApJ...772L..24R} {772, L24}

\bibitem[\protect\citeauthoryear{{Scaringi}, {Groot}, {Knigge}, {Lasota}, {de Martino}, {Cavecchi}, {Buckley}  \& {Camisassa}}{{Scaringi} et~al.}{2022a}]{2022MNRAS.514L..11S}
{Scaringi} S.,  {Groot} P.~J.,  {Knigge} C.,  {Lasota} J.~P.,  {de Martino} D.,  {Cavecchi} Y.,  {Buckley} D.~A.~H.,   {Camisassa} M.~E.,  2022a, \mn@doi [\mnras] {10.1093/mnrasl/slac042}, \href {https://ui.adsabs.harvard.edu/abs/2022MNRAS.514L..11S} {514, L11}

\bibitem[\protect\citeauthoryear{{Scaringi} et~al.,}{{Scaringi} et~al.}{2022b}]{2022Natur.604..447S}
{Scaringi} S.,  et~al., 2022b, \mn@doi [\nat] {10.1038/s41586-022-04495-6}, \href {https://ui.adsabs.harvard.edu/abs/2022Natur.604..447S} {604, 447}

\bibitem[\protect\citeauthoryear{{Schmidt}, {West}, {Liebert}, {Green}  \& {Stockman}}{{Schmidt} et~al.}{1986}]{1986ApJ...309..218S}
{Schmidt} G.~D.,  {West} S.~C.,  {Liebert} J.,  {Green} R.~F.,   {Stockman} H.~S.,  1986, \mn@doi [\apj] {10.1086/164593}, \href {http://adsabs.harvard.edu/abs/1986ApJ...309..218S} {309, 218}

\bibitem[\protect\citeauthoryear{Shapiro \& Teukolsky}{Shapiro \& Teukolsky}{1983}]{shapiro/2008}
Shapiro S.~L.,  Teukolsky S.~A.,  1983, Black holes, white dwarfs and neutron stars: the physics of compact objects.
John Wiley \& Sons

\bibitem[\protect\citeauthoryear{{Shara}}{{Shara}}{1982}]{1982ApJ...261..649S}
{Shara} M.~M.,  1982, \mn@doi [\apj] {10.1086/160376}, \href {https://ui.adsabs.harvard.edu/abs/1982ApJ...261..649S} {261, 649}

\bibitem[\protect\citeauthoryear{{Sousa}, {Coelho}  \& {de Araujo}}{{Sousa} et~al.}{2020}]{2020MNRAS.492.5949S}
{Sousa} M.~F.,  {Coelho} J.~G.,   {de Araujo} J. C.~N.,  2020, \mn@doi [\mnras] {10.1093/mnras/staa205}, \href {https://ui.adsabs.harvard.edu/abs/2020MNRAS.492.5949S} {492, 5949}

\bibitem[\protect\citeauthoryear{{Sousa}, {Coelho}, {de Araujo}, {Kepler}  \& {Rueda}}{{Sousa} et~al.}{2022}]{2022ApJ...941...28S}
{Sousa} M.~F.,  {Coelho} J.~G.,  {de Araujo} J.~C.~N.,  {Kepler} S.~O.,   {Rueda} J.~A.,  2022, \mn@doi [\apj] {10.3847/1538-4357/aca015}, \href {https://ui.adsabs.harvard.edu/abs/2022ApJ...941...28S} {941, 28}

\bibitem[\protect\citeauthoryear{{Sousa}, {Coelho}, {de Araujo}, {Guidorzi}  \& {Rueda}}{{Sousa} et~al.}{2023}]{2023ApJ...958..134S}
{Sousa} M.~F.,  {Coelho} J.~G.,  {de Araujo} J.~C.~N.,  {Guidorzi} C.,   {Rueda} J.~A.,  2023, \mn@doi [\apj] {10.3847/1538-4357/ad022f}, \href {https://ui.adsabs.harvard.edu/abs/2023ApJ...958..134S} {958, 134}

\bibitem[\protect\citeauthoryear{{Terada} et~al.,}{{Terada} et~al.}{2008}]{Terada2008}
{Terada} Y.,  et~al., 2008, \mn@doi [\pasj] {10.1093/pasj/60.2.387}, \href {https://ui.adsabs.harvard.edu/abs/2008PASJ...60..387T} {60, 387}

\bibitem[\protect\citeauthoryear{{The LIGO Scientific Collaboration}, {the Virgo Collaboration}, {the KAGRA Collaboration}, {Abbott}, {Abbott}  et~al.}{{The LIGO Scientific Collaboration} et~al.}{2021}]{2021arXiv211103606T}
{The LIGO Scientific Collaboration} {the Virgo Collaboration} {the KAGRA Collaboration} {Abbott} R.,  {Abbott} T.~D.,   et~al., 2021, \mn@doi [arXiv e-prints] {10.48550/arXiv.2111.03606}, \href {https://ui.adsabs.harvard.edu/abs/2021arXiv211103606T} {p. arXiv:2111.03606}

\bibitem[\protect\citeauthoryear{{Usov}}{{Usov}}{1988}]{1988PAZh...14..606U}
{Usov} V.~V.,  1988, Pisma v Astronomicheskii Zhurnal, \href {https://ui.adsabs.harvard.edu/abs/1988PAZh...14..606U} {14, 606}

\bibitem[\protect\citeauthoryear{{Warner}}{{Warner}}{2003}]{2003cvs..book.....W}
{Warner} B.,  2003, {Cataclysmic Variable Stars}.
Cambridge Astrophysics, Cambridge University Press, \mn@doi{10.1017/CBO9780511586491}

\bibitem[\protect\citeauthoryear{Weber}{Weber}{1999}]{weber_pulsars_1999}
Weber F.,  1999, Pulsars as {Astrophysical} {Laboratories} for {Nuclear} and {Particle} {Physics}, 1st edition edn.
CRC Press, Bristol ; Philadelphia

\bibitem[\protect\citeauthoryear{{Weber} \& {Glendenning}}{{Weber} \& {Glendenning}}{1992}]{1992ApJ...390..541W}
{Weber} F.,  {Glendenning} N.~K.,  1992, \mn@doi [\apj] {10.1086/171304}, \href {http://adsabs.harvard.edu/abs/1992ApJ...390..541W} {390, 541}

\bibitem[\protect\citeauthoryear{{Welsh}, {Horne}  \& {Gomer}}{{Welsh} et~al.}{1998}]{1998MNRAS.298..285W}
{Welsh} W.~F.,  {Horne} K.,   {Gomer} R.,  1998, \mn@doi [\mnras] {10.1046/j.1365-8711.1998.01643.x}, \href {https://ui.adsabs.harvard.edu/abs/1998MNRAS.298..285W} {298, 285}

\bibitem[\protect\citeauthoryear{{Yagi} \& {Seto}}{{Yagi} \& {Seto}}{2017}]{2017PhRvD..95j9901Y}
{Yagi} K.,  {Seto} N.,  2017, \mn@doi [\prd] {10.1103/PhysRevD.95.109901}, \href {https://ui.adsabs.harvard.edu/abs/2017PhRvD..95j9901Y} {95, 109901}

\bibitem[\protect\citeauthoryear{{Yakovlev}, {Gasques}, {Afanasjev}, {Beard}  \& {Wiescher}}{{Yakovlev} et~al.}{2006}]{2006PhRvC..74c5803Y}
{Yakovlev} D.~G.,  {Gasques} L.~R.,  {Afanasjev} A.~V.,  {Beard} M.,   {Wiescher} M.,  2006, \mn@doi [\prc] {10.1103/PhysRevC.74.035803}, \href {https://ui.adsabs.harvard.edu/abs/2006PhRvC..74c5803Y} {74, 035803}

\bibitem[\protect\citeauthoryear{{de Araujo}, {Coelho}  \& {Costa}}{{de Araujo} et~al.}{2016}]{2016ApJ...831...35D}
{de Araujo} J. C.~N.,  {Coelho} J.~G.,   {Costa} C.~A.,  2016, \mn@doi [\apj] {10.3847/0004-637X/831/1/35}, \href {https://ui.adsabs.harvard.edu/abs/2016ApJ...831...35D} {831, 35}

\bibitem[\protect\citeauthoryear{{de Araujo}, {Coelho}  \& {Costa}}{{de Araujo} et~al.}{2017}]{2017EPJC...77..350D}
{de Araujo} J. C.~N.,  {Coelho} J.~G.,   {Costa} C.~A.,  2017, \mn@doi [European Physical Journal C] {10.1140/epjc/s10052-017-4925-3}, \href {https://ui.adsabs.harvard.edu/abs/2017EPJC...77..350D} {77, 350}

\makeatother
\end{thebibliography}

%\begin{thebibliography}{99}
%\end{thebibliography}

%%%%%%%%%%%%%%%%%%%%%%%%%%%%%%%%%%%%%%%%%%%%%%%%%%

%%%%%%%%%%%%%%%%% APPENDICES %%%%%%%%%%%%%%%%%%%%%

%\appendix



%%%%%%%%%%%%%%%%%%%%%%%%%%%%%%%%%%%%%%%%%%%%%%%%%%

% Don't change these lines
\bsp	% typesetting comment
\label{lastpage}
\end{document}